\documentclass[aps,prb,twocolumn,notitlepage,floatfix,nofootinbib,longbibliography]{revtex4-2}
\usepackage[latin9]{inputenc}
\usepackage{textcomp}
\usepackage{amsmath}
\usepackage{amssymb}
\usepackage{graphicx}

\makeatletter

\newcommand{\lyxmathsym}[1]{\ifmmode\begingroup\def\b@ld{bold}
  \text{\ifx\math@version\b@ld\bfseries\fi#1}\endgroup\else#1\fi}

\@ifundefined{textcolor}{}
{%
 \definecolor{BLACK}{gray}{0}
 \definecolor{WHITE}{gray}{1}
 \definecolor{RED}{rgb}{1,0,0}
 \definecolor{GREEN}{rgb}{0,1,0}
 \definecolor{BLUE}{rgb}{0,0,1}
 \definecolor{CYAN}{cmyk}{1,0,0,0}
 \definecolor{MAGENTA}{cmyk}{0,1,0,0}
 \definecolor{YELLOW}{cmyk}{0,0,1,0}
}

\usepackage{amsfonts}
\usepackage{amsmath}[fleqn]

\makeatother

\begin{document}
\title{Phenomenological theory of magnetic 90$^{\circ}$ helical state}
\author{A. E. Koshelev}
\affiliation{Materials Science Division, Argonne National Laboratory, Lemont, Illinois
60439}
\date{\today }
\begin{abstract}
We explore a phenomenological phase diagram for the magnetic helical
state with 90$^{\circ}$ turn angle between neighboring spins
in the external magnetic field. 
Such state is formed by the Eu spin layers in the superconducting iron arsenide  RbEuFe$_{4}$As$_{4}$.
The peculiarity of this spin configuration is that it is not realized in the standard Heisenberg model with bilinear
exchange interactions. A minimum model allowing for such a state requires
the biquadratic nearest-neighbor interaction term. In addition, in
tetragonal materials the 90$^\circ$ helix state
may be stabilized by the in-plane four-fold anisotropy term, which
also fixes helix orientation with respect to the crystal lattice. 
%Resub
Such a system has a very rich behavior in the external magnetic field.
%A key feature characterizing the behavior in the magnetic field is 
The magnetic field induces
the  metamagnetic transition to the double-periodic state with the moment angles  ($\alpha$,
$\alpha$, $-\alpha$, $-\alpha$) with respect to the field for the four subsequent spins. 
The transition field to this state from the
deformed helix is determined by the strength of biquadratic interaction. 
%When this  interaction is weak the double-periodic state occupies a wide field range.
The transition is second order for small biquadratic coupling and becomes first
order when this coupling exceeds the critical value. 
On the other hand, the aligned state at high magnetic field becomes unstable with respect to formation of incommensurate fan state which transforms into the double-periodic state with decreasing magnetic field. The range of this incommensurate state near the saturation field is proportional to square of the biquadratic coupling. 
In addition, when the magnetic field is applied along one of four the equilibrium
moment directions, the deformed helix state experience the first-order
rotation transition at the field determined by the four-fold anisotropy. 
\end{abstract}
\maketitle

\section{Introduction}

Several magnetic materials have noncollinear helical ground states,
in which spins ferromagnetically align within layers but rotate from
layer to layer at a finite angle around the helix axis. In materials
without inversion symmetry, such as MnSi\cite{IshikawaSSC76}, FeGe\cite{LebechFeGeJPhys1989},
and Fe$_{1\lyxmathsym{\textendash}x}$Co$_{x}$Si\cite{BeilleFeCoSuSSC83,UchidaSci06},
these helical configurations are caused by Dzyaloshinskii-Moriya interaction
and, due to the weakness of this interaction, the structure period is
very large. Helical states are also realized in materials with inversion
symmetry, such as EuNi$_{2}$As$_{2}$\cite{JinEuNi2As2PhysRevB.99.014425,SangeethaEuNi2As2PhysRevB.100.094438}
and EuCo$_{2}$P$_{2}$\cite{ReehuisEuCo2P2JPCS92,SangeethaEuCo2P2PhysRevB.94.014422}.
In this case, they emerge due to competing exchange interactions\cite{EnzJAP61,Nagamiya1968,JohnstonPhysRevB.91.064427,JohnstonPhysRevB.96.104405}.
The simplest case is the next-nearest neighbor classical Heisenberg
model described by the energy functional
\begin{align} E\!&=\!-\sum_{n}\left(J_{z,1}\boldsymbol{s}_{n}\boldsymbol{s}_{n+1}\!+\!J_{z,2}\boldsymbol{s}_{n}\boldsymbol{s}_{n+2}\right)
	\label{eq:HeisenbergModel}\\
&=\!-\sum_{n}\left[J_{z,1}\cos\left(\phi_{n+1}\!-\!\phi_{n}\right)\!+\!J_{z,2}\cos\left(\phi_{n+2}\!-\!\phi_{n}\right)\right]
\nonumber,
\end{align}	
where $\boldsymbol{s}_{n}\!=\!\left(\cos\phi_{n},\sin\phi_{n},0\right)$
is the unit in-plane vector along the spin direction in n$^{\mathrm{th}}$
layer. For this model, the ground helical state $\phi_{n}^{(0)}=Qn$
is realized for the relation between the exchange constants $J_{z,2}\!<\!-|J_{z,1}|/4$
with $\cos Q\!=\!-J_{z,1}/(4J_{z,2})$ \cite{EnzJAP61,Nagamiya1968,JohnstonPhysRevB.91.064427,JohnstonPhysRevB.96.104405}.
In metallic magnets, the frustration may be caused by the oscillating
interaction between spins mediated by conduction electrons, known
as Ruderman-Kittel-Kasuya-Yosida (RKKY) interaction. A special case
of helical magnetism is predicted in superconducting ferromagnets
\cite{BulaevskiiJLTP1980,BulaevskiiAdvPhys85,AbrikosovBook,KulicBuzdinSupercondBook},
where the origin of frustration is the competition between the normal
and superconducting RKKY interactions.  Experimentally, a helical magnetic
state has been established in the nickel borocarbide HoNi$_{2}$B$_{2}$C\cite{MullerRoPP2001,GuptaAdvPhys06}.
Such state may also realize in superconducting compound
ErRh$_{4}$B$_{4}$\cite{BulaevskiiJLTP1980,WolowiecPhysC15}.
%May be domain structure?

Recently, the helical magnetic state has been found in the superconducting
iron pnictide RbEuFe$_{4}$As$_{4}$ \cite{IidaPhysRevB.100.014506,IslamPreprint2019}.
In this state, the Eu moments align ferromagnetically in Eu layers and rotate 90$^\circ$ from layer to layer. Such 90$^\circ$ helix state is unique, because it does not exist within the above simple Heisenberg exchange model \cite{JohnstonPhysRevB.91.064427} 
and its stabilization requires spin interactions beyond common bilinear terms. 
Indeed, in the above framework, formally, the 90$^\circ$ helix, $Q\!=\!\pi/2$, is
supposed to realize when the nearest-neighbor constant vanishes, $J_{z,1}\!=\!0$, and the next-nearest-neighbor constant is negative, $J_{z,2}<0$.
However, in this case, the interaction between two sublattices composed
of odd and even spin layers is absent so that the energy is degenerate with
respect to relative rotation of these sublattices\cite{NagamiyaPTP61,JohnstonPhysRevB.96.104405}.
The 90$^\circ$ helix is only one of such states
corresponding to the orthogonal orientation of the sublattices moments.
Adding interactions with more remote layers does not resolve this
issue. Therefore, the stabilization of the 90$^\circ$
helical state requires inclusions of nonconventional spin interactions.
The simplest such interaction is the nearest-neighbor biquadratic
term $J_{z,b}\left(\boldsymbol{s}_{n}\boldsymbol{s}_{n+1}\right)^{2}$.
It breaks rotational degeneracy between two sublattices and, when
the interaction constant is positive, $J_{z,b}>0$, favors 90$^\circ$
helix. 

%Resub
Biquadratic spin interactions between local magnetic moments were considered before in different
situations. Such interaction was first introduced in Ref.~\cite{HarrisPhysRevLett63}
to describe the interaction between Mn$^{2+}$ moments inside MgO crystal.
Later, the presence of the biquadratic interaction has been experimentally
established in the Fe-Cr-Fe sandwiches \cite{RuhrigPSSA91,DemokritovJPhysD1998}.
In particular, for certain thicknesses of interlayer Cr, this interaction
leads to the perpendicular orientation of the magnetization in two Fe
films. After this discovery, the presence of the biquadratic interaction
has been demonstrated in many other sandwich structures, see review
\cite{DemokritovJPhysD1998} and references therein. %
More recently, the biquadratic term was introduced to explain unusual 'up-up-down-down'
magnetic structure in several manganites, such as HoMnO$_{3}$\cite{KaplanPhysRevB.80.012407}.
Contrary to 90$^\circ$ helix, such double-periodic state is realized for \emph{negative} $J_{z,b}$.
%Resub
The most straightforward intrinsic origin of the biquadratic coupling between magnetic moments in metallic systems is 
%Microscopically, the biquadratic coupling in metallic systems originates from 
the higher-order expansion terms of the energy with respect to the exchange interaction between the moments and conduction electrons, see, e.g., Refs. \cite{BrunoPhysRevB.52.411,HayamiPhysRevB.95.224424}. %
Other mechanisms for this coupling were also theoretically proposed for different physical systems \cite{KaplanPhysRevB.80.012407,SlonczewskiJAP93,BastardisPhysRevB.76.132412}.

Biquadratic interactions are also likely are relevant for the magnetic
properties of iron pnictides. For example, the biquadratic coupling between the
Fe spins has to be taken into account to explain the domain-wall
structure and the spin-wave spectrum in the stripe antiferromagnetic state in FeAs layers\cite{WysockiNPhys2011}. 
The biquadratic coupling between Eu and Fe spins in EuFe$_{2}$As$_{2}$
has been also introduced in Ref.~\cite{MaiwaldPhysRevX.8.011011,SanchezPhysRevB.104.104413}
to model magnetic detwinning. Therefore, the assumption of a noticeable biquadratic interaction between Eu spin layers in  RbEuFe$_{4}$As$_{4}$ does not look too exotic. The vanishing of nearest-neighbor exchange in this material may be the result of an accidental compensation of the normal and superconducting contributions to the RKKY interaction \cite{KoshelevPhysRevB19}. The  biquadratic term is most likely caused by the interaction between spins mediated by superconducting electrons.
%Resub
\footnote{One can expect that the superconducting subsystem enhances the higher-order interactions between the local moments, 
	since Cooper pairing is sensitive to the exchange field. For the superconducting energy, 
	the expansion parameter is the ratio exchange field/superconducting gap. }

In the model with only bilinear and biquadratic exchange interactions,
the energy is degenerate with respect to helix rotation. This continuous
degeneracy is eliminated by the 4-fold crystal anisotropy term, $-K_{4}(s_{x,\boldsymbol{i},n}^{4}+s_{y,\boldsymbol{i},n}^{4})$.
In addition, such anisotropy term locks $Q\!=\!\pi/2$ state within
a finite range of the small next-neighbor exchange constant $J_{z,1}$, 
%Resub
see Appendix \ref{app-rangeJz1}. 
Nevertheless,
since the 90$^\circ$ helix only exists if $J_{z,1}$ is small, 
%in comparison with $K_4$, 
for simplicity,  we  assume that $J_{z,1}\!=\!0$ in the main text. Therefore, the 90$^\circ$ helical state in the magnetic field applied perpendicular to the helix axis can be described by the following energy functional
\begin{widetext}
\begin{align}
\mathcal{E} & =\sum_{n}\left[|J_{z,2}|\boldsymbol{s}_{n}\boldsymbol{s}_{n+2}+J_{z,b}\left(\boldsymbol{s}_{n}\boldsymbol{s}_{n+1}\right)^{2}+K_{4}\left(s_{x,n}^{4}+s_{y,n}^{4}-\frac{3}{4}\right)-\mu\boldsymbol{H}\boldsymbol{s}_{n}\right]\nonumber \\
=\! & \sum_{n}\left[|J_{z,2}|\cos\left(\phi_{n+2}\!-\!\phi_{n}\right)\!+\!J_{z,b}\cos^{2}\left(\phi_{n+1}\!-\!\phi_{n}\right)
%\!+\!K_{4}\left(\cos^{4}\phi_{n}\!+\!\sin^{4}\phi_{n}\!-\frac{3}{4}\right)
\!+\!\frac{K_{4}}{4}\cos(4\phi_{n})
\!-\!\mu H\cos\left(\phi_{n}\!-\!\theta\right)\right],
\label{eq:InterLayEnerHK4}
\end{align}
\end{widetext}
where $\mu=g\mu_{B}S$ is the magnetic moment, $S$
is the spin, and $\theta$ is the in-plane angle of the external magnetic
field $\boldsymbol{H}\!=\!H(\cos \theta, \sin \theta, 0)$. At zero field the ground state of this model is given by the 90$^\circ$ helix with one of the spins oriented at 45$^\circ$ with respect to the $x$
axis. As other spirals, the ground state is chiral and degenerate with respect
to the direction of rotation, i.e., the spiral can be either right hand or left hand.

The goal of this paper is to investigate the magnetic phase diagram
following from the energy functional in Eq.~\eqref{eq:InterLayEnerHK4}.
The overall behavior for this model is fully determined by the two
dimensionless parameters, $r_{b}=J_{z,b}/|J_{z,2}|$ and $k_{4}\equiv K_{4}/|J_{z,2}|$,
both of which are expected to be small. We also introduce the reduced
magnetic field $\tilde{h}=\mu H/|J_{z,2}|$. The equilibrium spin
angles $\phi_{n}$ determine the magnetization per spin, $m=\left\langle \cos\phi_{n}\right\rangle $.
In real units it determines the bulk magnetization normalized to its saturation
value, $m=M/M_{\mathrm{sat}}$, with $M_{\mathrm{sat}}=\mu n_{M}$,
where $n_{M}$ is the bulk density of the moments.

A detailed investigation of helical magnetic structures with
different commensurate modulation wave vectors $Q$ within the Heisenberg model in Eq.\ \eqref{eq:HeisenbergModel} has been performed
in Ref.~\cite{JohnstonPhysRevB.96.104405}. A small magnetic field
applied perpendicular to the helix axis distorts the helix. With further
increase of the field, the distorted helix transforms into the fanlike
structure with $\phi_{n}=\phi_{\mathrm{max}}\sin[(n+\alpha)Q]$ first proposed in Ref.\ \cite{EnzJAP61}.
The nature of this transformation is determined by the modulation
wave vectors $Q$ and several different scenarios may be realized\cite{JohnstonPhysRevB.96.104405}. 
For long-wave structures with $Q<4\pi/9$, the first-order transition takes place at a certain transition field $H_{t}$, at which the magnetization jumps. 
In the range $4\pi/9<Q<\pi$ the behavior is not universal.
Typically, the transformation occurs as a smooth crossover but at
several commensurate values of $Q$ either first- or second-order
phase transitions take place. The fan state exists until the magnetic
field reaches the saturation field, at which the fan angle $\phi_{\mathrm{max}}$ vanishes. 

%Resub
We will see that the phase diagram of 90$^\circ$ helix following from Eq.~\eqref{eq:InterLayEnerHK4} is very rich and has both similarities with and differences from other magnetic spiral structures. The main findings of this paper can be summarized as follows. (i)Small magnetic field distorts the helix and induces phase transition into the double-periodic state, in which the four subsequent spins form the  angles  ($\alpha$, $\alpha$, $-\alpha$, $-\alpha$) with respect to the field. This state is actually a particular case of a commensurate fan\cite{JohnstonPhysRevB.96.104405}. At small biquadratic coupling this transition is continuous and it becomes first order when the biquadratic coupling exceeds a certain critical value. (ii) The saturated state at high magnetic fields becomes unstable with respect to the incommensurate-fan state with decreasing field. This state occupies the field range proportional to the biquadratic coupling squared and transforms into the double-periodic state when magnetic field decreases below certain level. (iii)For the finite in-plane four-fold anisotropy and the magnetic field applied along one of the four easy-axis directions, there is an additional small-field first-order phase transition corresponding to 45$^\circ$ rotation of the helix.
\begin{figure*}
	\includegraphics[width=6.6in]{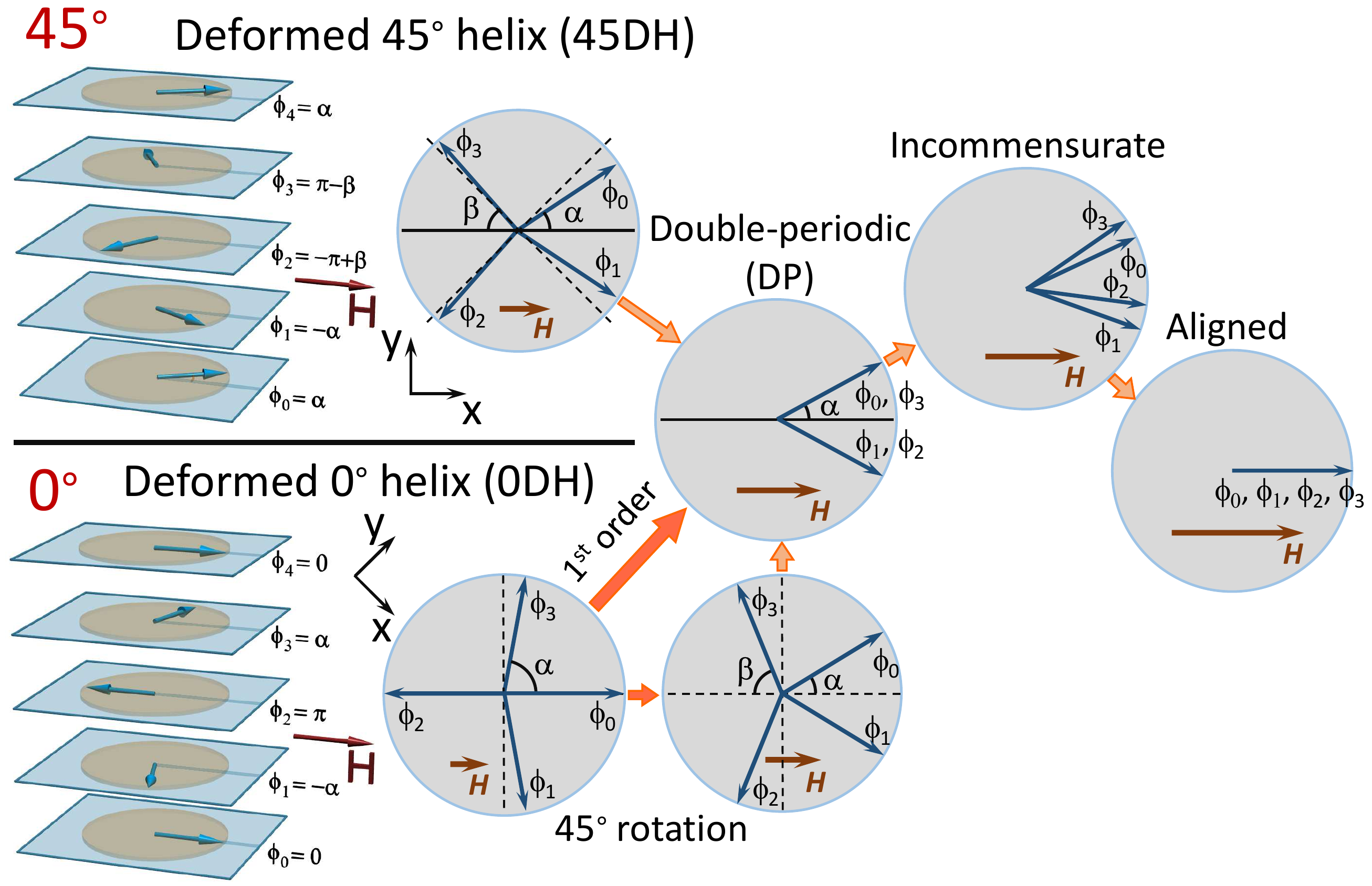}
	\caption{Spin configurations realized with increasing magnetic field for two
		field orientations for the model with four-fold anisotropy. The upper sequence
		also describes states for the rotationally-degenerate model with $k_{4}=0$. }
	\label{Fig:SpinStatesK4}
\end{figure*}

The paper is organized as follows. In Sec.~\ref{sec:RotDegen}, we
investigate the phase diagram of the rotationally-degenerate model
with $k_{4}\!=\!0$, which is determined by only one parameter $r_{b}$. We find the transition field to the double-periodic state as a function of this parameter and evaluate the value of $r_{b}$ above which the transition becomes first order. 
%Resub
Analyzing the same system at high magnetic fields, we find that the incommensurate-fan state emerges below the saturation field. We evaluate the transition field between this state and the double-periodic state. %
We also compute the field dependences of the magnetization for different $r_{b}$ and find its features at the transition points.
In Sec.~\ref{sec:FiniteK4}, we study the model with finite
four-fold anisotropy which fixes helix orientation and sets four easy-axis directions. We investigate the phase diagrams and the magnetization for the magnetic field applied along the two symmetry directions, along the equilibrium spin orientation and at 45$^\circ$ with respect to this direction. In particular, for the former field orientation, we investigate the first-order helix-rotation transition. Finally, we summarize and discuss the obtained results in Sec.~\ref{sec:summary}.

\section{Rotationally-degenerate system without four-fold anisotropy\label{sec:RotDegen}}

Before the investigation of the full model described by Eq.\ \eqref{eq:InterLayEnerHK4}, we consider a simpler model without four-fold anisotropy, $K_{4}=0$, corresponding to a rotationally-degenerate helix. In this case, the magnetic response is controlled by one reduced parameter $r_b$. In general, the equilibrium configuration of the spin angles $\phi_{n}$ obeys the equations
\begin{align}
 & \sin\left(\phi_{n+2}\!-\!\phi_{n}\right)\!+\!\sin\left(\phi_{n-2}\!-\!\phi_{n}\right)\nonumber \\
&+  r_{b}\left(\sin\left[2\left(\phi_{n+1}\!-\!\phi_{n}\right)\right]\!+\!\sin\left[2\left(\phi_{n-1}\!-\!\phi_{n}\right)\right]\right)\nonumber \\
&+  \tilde{h}\sin\left(\phi_{n}\!-\!\theta\right)=\!0.\label{eq:EqPhasesGeneral}
\end{align}
If we assume that the four-spin periodicity is maintained in the magnetic
field, $\phi_{n+4}=\phi_{n}$ then Eq.~\eqref{eq:EqPhasesGeneral}
gives four nonlinear equations for the four phases with $n=0,1,2,3$. 
%Resub
Further analysis shows that for small $r_b$ the four-spin periodicity is maintained in the most part of the phase diagram but is violated near the saturation field.

The magnetic field breaks down the rotational degeneracy of the helix. %
%Resub
In the energy expansion with respect to the magnetic field $\tilde{h}$, the quadratic term is isotropic with respect to the field orientation, while the quartic term, in addition to  an isotropic contribution, has also the angle-dependent part proportional to $\tilde{h}^4(\cos^4\vartheta_h\! +\! \sin^4\vartheta_h)$, where $\vartheta_h$ is the angle between the field and one of the equilibrium moment directions. The favorable helix orientation is determined by the sign of the coefficient in front of this term. %
To find this equilibrium helix orientation at a small magnetic field,
we compare the energies for two symmetric orientations shown in Fig.~\ref{Fig:SpinStatesK4},
which we refer to as deformed 45$^\circ$ helix (45DH) and deformed 0$^\circ$ helix (0DH) shown
in the upper and lower 3D picture, respectively. The energy difference between these orientations %
%Resub
%appears only in the quartic order with respect to $\tilde{h}$.
determines the coefficient in the above angle-dependent quartic term in the energy expansion. 

For the 45$^\circ$ helix, the four angles $\phi_{n}$
are determined by only two independent angles $\alpha$ and $\beta$,
see the left picture in the upper part of Fig.\ \ref{Fig:SpinStatesK4},
\begin{equation}
\phi_{0}=\alpha,\,\phi_{1}=-\alpha,\,\phi_{2}=\pi+\beta,\,\phi_{3}=\pi-\beta,\label{eq:SymState}
\end{equation}
and at zero field, we have $\alpha\!=\!\beta\!=\!\pi/4$. The reduced energy
per spin, $E_{s}=\mathcal{E}_{\mathrm{lay}}/N|J_{z,2}|$, for this
state 
\begin{align}
E_{s}\left(\alpha,\beta\right) & \!=\!-\cos\left(\beta\!-\!\alpha\right)\!\nonumber \\
+ & \frac{r_{b}}{4}\left[\cos^{2}\left(2\alpha\right)\!+\!\cos^{2}\left(2\beta\right)\!+\!2\cos^{2}\left(\beta\!+\!\alpha\right)\right]\nonumber \\
- & \frac{\tilde{h}}{2}\left(\cos\alpha-\cos\beta\right)
\label{eq:EnSymState}
\end{align}
yields the equations for the equilibrium angles $\alpha$ and $\beta$
\begin{subequations}
%{\setlength{\mathindent}{0cm}
\begin{align}
&2\sin\left(\beta\!-\!\alpha\right)\!+\!r_{b}\left(\sin(4\alpha)\!+\!\sin\left[2(\alpha\!+\!\beta)\right]\right)\!-\!\tilde{h}\sin\alpha  \!=\!0,\ \ \ \ \label{eq:alpha-beta}\\
&2\sin\left(\beta\!-\!\alpha\right)\!-\!r_{b}\left(\sin(4\beta)\!+\!\sin\left[2(\alpha\!+\!\beta)\right]\right)\!-\!\tilde{h}\sin\beta  \!=\!0.\ \ \ \ 
\end{align}
%}
\end{subequations}
These two angles determine the reduced magnetization
per spin, 
\begin{equation}
m=\left(\cos\alpha-\cos\beta\right)/2.\label{eq:MagnSymState}
\end{equation}

Small magnetic field induces small deviations, which we represent
as $\alpha\!=\!\frac{\pi}{4}\!-\!\alpha_{+}\!-\frac{\alpha_{-}}{2}$,
$\beta\!=\!\frac{\pi}{4}+\!\alpha_{+}\!-\frac{\alpha_{-}}{2}$. Expanding
the energy with respect to the small deviations, $\alpha_{\pm}$,
we obtain
\begin{align}
E_{s}\left(\alpha_{+},\alpha_{-}\right) & \approx-1+2\left(1+r_{b}\right)\alpha_{+}^{2}+r_{b}\alpha_{-}^{2}\nonumber \\
- & \frac{2(1\!+\!4r_{b})}{3}\alpha_{+}^{4}-\frac{r_{b}}{3}\left(12\alpha_{+}^{2}\alpha_{-}^{2}+\alpha_{-}^{4}\right)\nonumber \\
- & \frac{\sqrt{2}\tilde{h}}{4}\left(2\alpha_{+}\!-\!\alpha_{+}\alpha_{-}\!-\frac{1}{3}\alpha_{+}^{3}\!-\frac{1}{4}\alpha_{+}\alpha_{-}^{2}\right).\label{eq:EnSymStateExpan-1}
\end{align}
Minimization of this energy gives equations for equilibrium $\alpha_{\pm}$,
\begin{subequations}
\begin{align}
 & 4\left(1\!+\!r_{b}\right)\alpha_{+}\!-\frac{8(1\!+\!4r_{b})}{3}\alpha_{+}^{3}\!-\!8r_{b}\alpha_{+}\alpha_{-}^{2}\!\nonumber \\
 & -\frac{\sqrt{2}\tilde{h}}{4}\left(2\!-\!\alpha_{-}\!-\!\alpha_{+}^{2}\!-\!\frac{1}{4}\alpha_{-}^{2}\right)=0,\\
 & 2r_{b}\alpha_{-}\!-\frac{4r_{b}}{3}\left(6\alpha_{+}^{2}\alpha_{-}\!+\!\alpha_{-}^{3}\right)\!+\frac{\sqrt{2}\tilde{h}}{4}\left(\alpha_{+}\!+\frac{1}{2}\alpha_{+}\alpha_{-}\right)\!=\!0,\label{eq:Equil-alp-pm}
\end{align}
\end{subequations}
which we expand with respect to $\tilde{h}$
\begin{equation}
\alpha_{\pm}=\sum_{n=1}^{\infty}\alpha_{\pm}^{(n)}\tilde{h}^{n}.\label{eq:alpha-h-exp}
\end{equation}
A simple analysis shows that only odd $\alpha_{+}^{(n)}$ and even $\alpha_{-}^{(n)}$
are finite, $\alpha_{+}^{(2k)}=\alpha_{-}^{(2k-1)}=0$. For the expansion
coefficients, we derive from Eqs.~\eqref{eq:Equil-alp-pm}\begin{subequations}
\begin{align}
\alpha_{+}^{(1)} & =\frac{\sqrt{2}}{8\left(1+r_{b}\right)},\label{eq:alp-pm-linear}\\
\alpha_{-}^{(2)} & =-\frac{1}{32r_{b}\left(1+r_{b}\right)},\label{eq:alp-pm-2}\\
\alpha_{+}^{(3)} & =\frac{\sqrt{2}}{512\left(1\!+\!r_{b}\right)^{4}}\left[\frac{4(1\!+\!4r_{b})}{3}+\!1\!+\frac{1}{r_{b}}\right],\label{eq:alp-pm-3}
\end{align}
\end{subequations}Substituting the expansion in Eq.~\eqref{eq:alpha-h-exp}
with these coefficients into Eq.~\eqref{eq:EnSymStateExpan-1}, we
finally obtain the energy expansion
\begin{equation}
E_{s}(\tilde{h})\!\approx\!-1\!-\frac{\tilde{h}^{2}}{16\left(1\!+\!r_{b}\right)}-\frac{1\!+\!2r_{b}^{2}/\left(1\!+\!r_{b}\right)^{2}}{1024r_{b}\left(1\!+\!r_{b}\right)^{2}}\tilde{h}^{4}.\label{eq:SymEs-h-exp}
\end{equation}
From this result, we can also obtain the magnetization at small fields,
$m=-dE_{s}/d\tilde{h}$,
\begin{equation}
m(\tilde{h})\approx\frac{\tilde{h}}{8\left(1\!+\!r_{b}\right)}+\frac{1\!+\!2r_{b}^{2}/\left(1\!+\!r_{b}\right)^{2}}{256r_{b}\left(1\!+\!r_{b}\right)^{2}}\tilde{h}^{3}.\label{eq:MagnK40SmallH}
\end{equation}
It is characterized by the upward curvature which becomes more pronounced
at smaller $r_{b}$.

Consider now the deformed 0$^\circ$ helix, see Fig.~\ref{Fig:SpinStatesK4}. This
state is determined by just one angle $\alpha$ as $\phi_{0}=0$,
$\phi_{3}=-\phi_{1}=\alpha$, $\phi_{2}=\pi$, and its energy is
\begin{align}
E_{s} & \!=-1+\left(1+r_{b}\right)\cos^{2}\alpha-\frac{\tilde{h}}{2}\cos\alpha.\label{eq:En0DegState}
\end{align}
Finding its minimum, we obtain
\begin{subequations}
\begin{align}
\cos\alpha & =\frac{\tilde{h}}{4\left(1+r_{b}\right)},\label{eq:AlignAlpha} \\
E_{s}(\tilde{h}) & =-1-\frac{\tilde{h}^{2}}{16\left(1+r_{b}\right)}.\label{eq:AlignEs-h-exp}
\end{align}
\end{subequations}
Note that the quartic term is absent for this helix orientation. Comparing
this result with Eq.~\eqref{eq:SymEs-h-exp}, %
%Resub
we find the quartic angle-dependent term in the energy 
\begin{align}
E_{s}^{(4)}(\vartheta_h)&=C_{4}\tilde{h}^{4}\left[-1+\cos^{4}\vartheta_h\!+\!\sin^{4}\vartheta_h\right]\nonumber\\
&=-2C_{4}\tilde{h}^{4}\cos^{2}\vartheta_h\sin^{2}\vartheta_h\\
&\text{with }\ C_{4}=\frac{1+2r_{b}^{2}/\left(1+r_{b}\right)^{2}}{512r_{b}\left(1+r_{b}\right)^{2}}\nonumber
\label{eq:QuartAngleTerm}
\end{align}
We see that the magnetic field breaks down the rotational degeneracy and favors the 45$^\circ$
deformed helix state shown in Fig.~\ref{Fig:SpinStatesK4}. From
now on, we consider the evolution of this state with increasing magnetic
field.

A finite biquadratic coupling favors perpendicular orientation between
the moments in odd and even layers. On the other hand, the magnetic
field wants to orient the moments in both sublattices in the same
way. This means that at small $r_{b}$, one can expect a metamagnetic transition
to the double-periodic (DP) state with increasing the magnetic field
with $\phi_{3}=\phi_{0}$ and $\phi_{2}=\phi_{1}$, (the state in the middle of
Fig.~\ref{Fig:SpinStatesK4}). Importantly, a
chiral nature of the state vanishes at this transition, since the mirror
image can be matched with the original spin configuration. We proceed
with the analysis of this double-periodic state and investigation
of the transition to it.

\subsection{Double-periodic state at high $\tilde{h}$}

At high fields the double-periodic state is realized, in which $\beta=\pi-\alpha$,
and
\begin{equation}
\phi_{0}=\phi_{3}=\alpha,\,\phi_{1}=\phi_{2}=-\alpha.\label{eq:DoublePerAngles}
\end{equation}
Since such angle distribution can also be presented as 
%Resub
$\phi_{n}=-\sqrt{2}\alpha\sin\left[(n\!-\!1/2)\pi/2\right]$,
this state formally belongs to the family of fanlike states\cite{JohnstonPhysRevB.96.104405}.
The energy of this double-periodic state 
\begin{align}
E_{s}\left(\alpha,\pi\!-\!\alpha\right) & \!=\!\cos\left(2\alpha\right)\!+\frac{r_{b}}{2}\left[1\!+\!\cos^{2}\left(2\alpha\right)\right]\!-\!\tilde{h}\cos\alpha\label{eq:EnDPerState}
\end{align}
gives the equation for the equilibrium angle
\begin{align}
4\cos\alpha_{0}\left[1+r_{b}\cos\left(2\alpha_{0}\right)\right] & =\tilde{h}.
\label{eq:DoublePerState}
\end{align}
The limit $\alpha_{0}\!=\!0$ formally corresponds to saturation (the right state
in Fig.~\ref{Fig:SpinStatesK4}). This condition
gives the nominal saturation field $\tilde{h}_{\mathrm{sat}}^{\mathrm{DP}}\!=4\left(1\! +\! r_{b}\right)$. We will show in the next section, however, that the double-periodic state transforms into a incommensurate-fan state with increasing magnetic field.  As a consequence, the true saturation field is somewhat higher than the above value.

To find the stability range of the double-periodic state, we consider
a small deviation from the double periodicity, $\alpha\!=\!\alpha_{0}\!+\!\psi$,
$\beta\!=\!\pi\!-\!\alpha_{0}\!+\!\psi$ with $\psi\!\ll\!1$. The energy expansion
\begin{align}
\delta E_{s}^{(2)}\left(\psi\right) & =\left[ -4r_{b}\cos^{2}\left(2\alpha_{0}\right)+\frac{\tilde{h}}{2}\cos\alpha_{0}\right] \psi^{2}\label{eq:DublePeriodQuad}
\end{align}
obtained from Eq.~\eqref{eq:EnSymState} gives the stability condition
\begin{equation}
-8r_{b}\cos^{2}\left(2\alpha_{0}\right)+\tilde{h}\cos\alpha_{0}>0.\label{eq:DoublePerStab}
\end{equation}
Combining this equation with Eq.~\eqref{eq:DoublePerState} for the
equilibrium angle, we find $\cos\left(2\alpha_{0}\right)$ at the
instability point
\begin{equation}
\cos\left(2\alpha_{0}\right)=-\frac{2}{1\!+\!r_{b}\!+\!\mathcal{R}_{0}}\label{eq:cos2a0}
\end{equation}
with 
$
\mathcal{R}_{0}\equiv\sqrt{\left(1\!+\!r_{b}\right)^{2}\!+\!12r_{b}}.
$
This gives
\begin{equation}
\cos\alpha_{0}=\frac{\sqrt{r_{b}}\left[r_{b}\!+\!7\!+\!\mathcal{R}_{0}\right]^{1/2}}{1+r_{b}+\mathcal{R}_{0}}.\label{eq:cosa0}
\end{equation}
For $r_{b}\rightarrow0$, the instability angle approaches $\pi/2$.
Eqs.\ \eqref{eq:cos2a0} and \eqref{eq:cosa0} allow us to obtain the analytic result for the instability
field from Eq.\ \eqref{eq:DoublePerStab}
\begin{equation}
\tilde{h}_{i}=\frac{4\sqrt{r_{b}}\left[r_{b}\!+\!7\!+\!\mathcal{R}_{0}\right]^{1/2}\!\left(1\!-\!r_{b}\!+\!\mathcal{R}_{0}\right)}{\left(1+r_{b}+\mathcal{R}_{0}\right)^{2}}.
\label{eq:InstField}
\end{equation}
The double-periodic state is stable for $\tilde{h}>\tilde{h}_{i}$. In
the limit $r_{b}\ll1$, we obtain a simple asymptotics
\begin{equation}
\tilde{h}_{i}\simeq4\sqrt{2r_{b}},\label{eq:InstFieldAsymp}
\end{equation}
i.~e., at small $r_{b}$ the instability field decrease proportionally
to $\sqrt{r_{b}}$ and in this case the double-periodic state occupies
a wide range of magnetic fields. In real units the result in Eq.~\eqref{eq:InstFieldAsymp}
becomes $H_{i}\simeq4\sqrt{2J_{z,b}|J_{z,2}|}/\mu$. 

The instability field in Eq.~\eqref{eq:InstField} corresponds to
a second-order phase transition only if the coefficient $c_{4}$ in
the quartic energy expansion term, $\delta E_{s}^{(4)}=\frac{1}{4}c_{4}\psi^{4}$, is positive. Calculations described in Appendix \ref{App:quartic-k40}
give the following result for the quartic coefficient at the instability
point,
\begin{align}
c_{4}(r_{b}) & =20r_{b}u^{2}-\frac{\left(1-u^{2}\right)\left(1+15r_{b}u\right)^{2}}{2\left(r_{b}+u\right)},\label{eq:DoublePer-c4}\\
u(r_{b})= & -\cos\left(2\alpha_{0}\right)=\frac{2}{1\!+\!r_{b}\!+\!\sqrt{\left(1\!+\!r_{b}\right)^{2}\!+\!12r_{b}}}.\nonumber 
\end{align}
The parameter $c_{4}(r_{b})$ is positive at small $r_{b}$ and becomes
negative at large $r_{b}$. This transition between the regimes takes
place at $c_{4}(r_{0})=0$, giving
\begin{equation}
r_{0}=\frac{5+\sqrt{70}}{135}\approx0.099\label{eq:r0}
\end{equation}
and $u(r_{0})=(19-2\sqrt{70})/3$. For $r_{b}>r_{0}$ the transition
between the 45DH and DP states becomes first order. In this case,
the first-order transition field exceeds the instability field in
Eq.~\eqref{eq:InstField}.

%Resub
\subsection{Incommensurate-fan instability near the saturation field \label{subs:IncommensK4eq0}}

In this subsection, we turn to the region of high magnetic fields
for the model with the rotational degeneracy and investigate the instability
of the aligned state. Namely, we consider the general fanlike state
\begin{equation}
	\phi_{n}=\vartheta\sin\left(qn+\varphi\right)
	\label{eq:GenFan}
\end{equation}
and find the minimum of energy with respect to the fan amplitude $\vartheta$,
wave vector $q$, and, possibly, phase shift $\varphi$. Substituting
this fan ansatz into the energy in Eq.~\eqref{eq:InterLayEnerHK4}
with $K_{4}\!=\!0$, we obtain the reduced energy per layer, $E_{\mathrm{fan}}\!=\!\mathcal{E}/N|J_{z,2}|$,
\begin{align}
	& E_{\mathrm{fan}}\left(\vartheta,q,\varphi\right)=\frac{1}{N}\sum_{n}\Big\{\cos\left[2\vartheta\sin q\cos\left(nq\!+\!\varphi\right)\right]\!\nonumber \\
	+ & \!\left.r\!_{b}\cos^{2}\!\left[\!2\vartheta\sin\frac{q}{2}\cos\!\left(nq\!+\!\frac{q}{2}\!+\!\varphi\right)\right]\!-\!\tilde{h}\cos\left[\vartheta\sin(nq\!+\!\varphi)\right]\!\right\}.
	\label{eq:EnGenFan}
\end{align}
Near the saturation, we can expand this energy with respect to the
fan amplitude $\vartheta$, 
\begin{equation}
	E_{\mathrm{fan}}\left(\vartheta,q,\varphi\right)\approx E_{\mathrm{fan}}\left(0,q,\varphi\right)+\frac{1}{2}a_{2}\left(q,\varphi\right)\vartheta^{2}+\frac{1}{4}a_{4}\left(q,\varphi\right)\vartheta^{4}.\label{eq:EnFanExpan}
\end{equation}
For the quadratic term, we obtain
\begin{align}
	& a_{2}\left(q,\varphi\right)=-2\sin^{2}q-2r_{b}(1-\cos q)+\frac{\tilde{h}}{2}\nonumber \\
	- & \!\left(2\sin^{2}q\!+\frac{\tilde{h}}{2}\right)\left\langle \cos\left(2nq\!+\!2\varphi\right)\right\rangle _{n}\nonumber \\
	- & 2r_{b}(1\!-\!\cos q)\left\langle \cos\left[\left(2n\!+\!1\right)q\!+\!2\varphi\right]\right\rangle _{n},\label{eq:Fan-Expan-a2Result}
\end{align}
where $\left\langle \ldots\right\rangle _{n}$ notates the averaging
over the layer index. The last two oscillating terms average to zero
unless $q=\pi$ yielding 
\begin{align}
	a_{2}\left(q\right) & =\!-2\sin^{2}q-2r_{b}(1\!-\!\cos q)+\frac{\tilde{h}}{2},\mathrm{\,for}\,q\neq\pi.\label{eq:FanExpan-a2}
\end{align}
Note that the quadratic coefficient does not depend on the phase shift
$\varphi$. The saturated state with $\vartheta=0$ is stable at given
$\tilde{h}$ if the coefficient $a_{2}\left(q\right)$ is positive
for all $q$. The instability first develops at the wave vector $q\!=\!Q$
where $a_{2}\left(q\right)$ is minimal. From Eq.~\eqref{eq:FanExpan-a2},
we immediately obtain
\begin{equation}
	\cos Q=-\frac{r_{b}}{2}\label{eq:OptQ}
\end{equation}
and $a_{2}\left(Q\right)=\!-2\left(1+\frac{r_{b}}{2}\right)^{2}+\frac{\tilde{h}}{2}$.
For small $r_{b}$, this corresponds to a weakly incommensurate state
with $Q$ slightly larger than $\pi/2$ and the period smaller than four
layers. The instability develops at the field 
\begin{equation}
	\tilde{h}_{\mathrm{sat}}\!=\!\left(r_{b}\!+\!2\right)^{2},\label{eq:hsat-ic}
\end{equation}
or, in real units,
$H_{\mathrm{sat}}\!=(2|J_{z,2}|+J_{b})^{2}/(\mu|J_{z,2}|)$. As expected, this field is larger than the nominal saturation field
for the double-periodic state $4\left(1\!+\!r_{b}\right)$ introduced after Eq.\ \eqref{eq:DoublePerState}, but the
difference is quadratic in $r_{b}^{2}$ and is very small for $r_b<0.5$, see Fig.\ \ref{Fig:PhDiagRotDeg}.

To find the energy and magnetization slightly below the instability
field, we also need the quartic term in the energy expansion, for
which the derivation similar to Eq.~\eqref{eq:Fan-Expan-a2Result}
yields
\begin{widetext}
	\begin{align*}
		& a_{4}\left(q,\varphi\right)=\sin^{4}q+8r_{b}\sin^{4}\frac{q}{2}-\frac{\tilde{h}}{16}\\
		+ & \frac{1}{3}\left(4\sin^{4}q\!+\frac{\tilde{h}}{4}\right)\left\langle \cos\left(2nq\!+\!2\varphi\right)\right\rangle _{n}\!+\frac{32}{3}r_{b}\sin^{4}\frac{q}{2}\left\langle \cos\left[\left(2n\!+\!1\right)q\!+\!2\varphi\right]\right\rangle _{n}\\
		+ & \frac{1}{6}\left(2\sin^{4}q\!-\frac{\tilde{h}}{8}\right)\left\langle \cos\left(4nq\!+\!4\varphi\right)\right\rangle _{n}\!+\frac{8}{3}r_{b}\sin^{4}\frac{q}{2}\left\langle \cos\left[2\left(2n\!+\!1\right)q\!+\!4\varphi\right]\right\rangle _{n}.
	\end{align*}
\end{widetext}
In this case, the oscillating terms average to zero
for $q\neq\frac{\pi}{2},\pi$. For such incommensurate states, we
obtain
\begin{equation}
	a_{4}\left(q\right)=\sin^{4}q\!+\!2r_{b}\left(1\!-\!\cos q\right)^{2}\!-\frac{\tilde{h}}{16},\,\mathrm{for}\,q\!\neq\!\frac{\pi}{2},\pi\label{eq:FanExpan-a4}
\end{equation}
and $a_{4}\left(Q\right)\!=\!\left(1\!+\!\frac{r_{b}}{2}\right)^{4}\!-\frac{\tilde{h}}{16}\!\approx\!\frac{3}{4}\left(1\!+\!\frac{r_{b}}{2}\right)^{4}$.
As the quadratic term in Eq.~\eqref{eq:FanExpan-a2}, the quartic
term for incommensurate states does not depend on the phase angle
$\varphi$. In contrast, the quartic term for the commensurate fan
with four-layer period
\begin{equation}
	a_{4}\left(\frac{\pi}{2},\varphi\right)\!=\!1\!+\!2r_{b}\!-\!\frac{\tilde{h}}{16}\!+\!\frac{1}{3}\left(1\!-\!2r_{b}\!-\!\frac{\tilde{h}}{16}\right)\cos\left(4\varphi\right)\label{eq:FanExpan-a4-90}
\end{equation}
does depend on $\varphi$. The value $\varphi$ giving the smallest
$a_{4}\left(\frac{\pi}{2},\varphi\right)$ is energetically favorable.
For $1\!-\!2r_{b}\!-\!\tilde{h}/16>0$ this minimum is realized at
$\varphi\!=\!\pi/4$ corresponding to the double-periodic state yielding  $a_{4}\left(\frac{\pi}{2},\frac{\pi}{4}\right)\!=\!\frac{1}{3}\left(2\!+\!8r_{b}\!-\!\frac{\tilde{h}}{8}\right)$.
Near the instability field for the commensurate state, $\tilde{h}\lesssim4(1\!+\!r_{b})$,
the above inequality is valid if $r_{b}<1/3$. It is important to
note that the quartic term for the four-layer-period state is smaller
than for the incommensurate state. 

Minimizing the energy of the incommensurate state with respect to
the amplitude 
\begin{align}
	\vartheta^{2} & =\frac{2\left[\sin^{2}q\!+\!r_{b}\left(1\!-\!\cos q\right)-\frac{\tilde{h}}{4}\right]}{\sin^{4}q+2r_{b}\left(1-\cos q\right)^{2}-\tilde{h}/16},\label{eq:AmplSatIncommens}
\end{align}
we obtain the energy
\begin{align}
	E_{\mathrm{fan}}\left(q\right) & \!\approx\!1\!+\!r_{b}\!-\!\tilde{h}-\frac{\left[\sin^{2}q\!+\!r_{b}\left(1\!-\!\cos q\right)\!-\frac{\tilde{h}}{4}\right]^{2}}{\sin^{4}q\!+\!2r_{b}\left(1\!-\!\cos q\right)^{2}\!-\!\tilde{h}/16}.\label{eq:EnSatIncommens}
\end{align}
Near the instability, we can neglect the field dependence of the wave
vector and set $q\!=\!Q$ yielding a simpler result
\begin{equation}
	E_{\mathrm{fan}}\left(Q\right)\approx1\!+\!r_{b}\!-\!\tilde{h}-\frac{\left[\left(1+\frac{r_{b}}{2}\right)^{2}\!-\frac{\tilde{h}}{4}\right]^{2}}{\left(1\!+\!\frac{r_{b}}{2}\right)^{4}-\frac{\tilde{h}}{16}}.\label{eq:EnSatIncommensQ}
\end{equation}
On the other hand, for the four-layer-period state with $q=\pi/2$,
we have
\begin{align}
	E_{\mathrm{fan}}\left(\frac{\pi}{2},\varphi\right) & \approx1\!+\!r_{b}\!-\!\tilde{h}\nonumber \\
	- & \frac{\left(1+r_{b}-\frac{\tilde{h}}{4}\right)^{2}}{1\!+\!2r_{b}\!-\!\frac{\tilde{h}}{16}+\frac{1}{3}\left[1\!-\!2r_{b}\!-\!\frac{\tilde{h}}{16}\right]\cos\left(4\varphi\right)}.\label{eq:EnSat4}
\end{align}
For $1\!-\!2r_{b}\!-\!\tilde{h}/16\!>\!0$, the ground state is at $\varphi\!=\!\pi/4$
with
\begin{equation}
	E_{\mathrm{fan}}\left(\frac{\pi}{2},\frac{\pi}{4}\right)=1+r_{b}-\tilde{h}-\frac{3\left(1+r_{b}-\frac{\tilde{h}}{4}\right)^{2}}{2+8r_{b}-\tilde{h}/8}.\label{eq:EnSat4-45}
\end{equation}

Even though the instability initially develops at the incommensurate
wave vector defined by Eq.~\eqref{eq:OptQ}, with further field decrease
the four-layer-period state wins due to the smaller quartic coefficient.
At small $r_{b}$ this transition takes place at field slightly smaller
than the instability field of the the double-periodic state, $4\left(1\!+\!r_{b}\right)$,
within the validity range of the small-amplitude expansion. Compare
energies in Eqs.~\eqref{eq:EnSatIncommensQ} and \eqref{eq:EnSat4-45},
we obtain the value of the transition field in the main order with
respect to $r_{b}\ll1$
\begin{align}
	\tilde{h}_{\mathrm{i-c}} & \approx4\left(1\!+\!r_{b}\right)\nonumber \\
	- & 2\left(\sqrt{\frac{3}{2}}\!+\!1\right)\left[1\!+\!4\left(1\!+\!\sqrt{\frac{2}{3}}\right)r_{b}\right]r_{b}^{2}.
	\label{eq:i-c-trans}
\end{align}
The incommensurate-fan state is realized above this field up to the
saturation field in Eq.~\eqref{eq:hsat-ic}. The field range of this
state decreases roughly proportional to $r_{b}^{2}$. 
This transition field is shown in Fig.\ \ref{Fig:PhDiagRotDeg} by dotted red line.
We can conclude that at small $r_b$ the most field range below the saturation field is occupied by the double-periodic state.  
Note that the transition field in Eq.~\eqref{eq:i-c-trans}
is approximate because it was obtained by the energy comparison assuming
a simple periodic fan state in Eq.~\eqref{eq:GenFan}. It is possible
that the emerging state is more complicated. The commensurate-incommensurate
transition typically takes place via formation of a periodic lattice
of solitons. Nevertheless, we expect that the range of such soliton-lattice
state is very narrow meaning that the accurate transition field should
be very close to the estimate in Eq.~\eqref{eq:i-c-trans}.

\subsection{Phase diagram and magnetization curves}
\begin{figure}
	\includegraphics[width=3.4in]{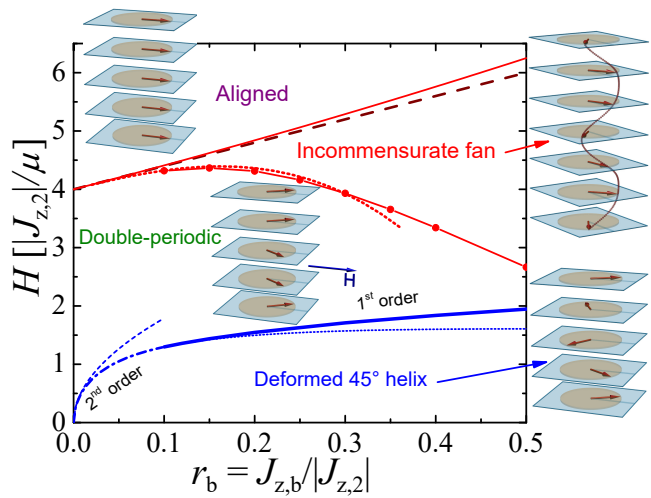}
	\caption{Magnetic phase diagram of the rotationally-isotropic model. The dotted
		blue line shows instability field of the double-periodic state, Eq.~\eqref{eq:InstField},
		which coincides with the second-order transition line for $r_{b}<0.099$
		(dash-dotted line). The dashed line shows the low-$r_{b}$ asymptotics
		of this line, Eq.~\eqref{eq:InstFieldAsymp}. 
		The blue solid line at higher $r_{b}$ shows the location of the first-order transition to the double-periodic state.
		The upper red solid line shows the transition between the saturated and incommensurate fan states. The red dotted line shows the approximate transition line from incommensurate to double-periodic state obtained from Eq.\ \eqref{eq:i-c-trans}. The solid red line with circles shows the more accurate transition line obtained by numerical minimization of the energy in Eq.\ \eqref{eq:EnGenFan} with respect to the fan state parameters. For comparison, the nominal saturation field for the double-periodic state is shown by the brown dashed line.}
	\label{Fig:PhDiagRotDeg}
\end{figure}
\begin{figure}
	\includegraphics[width=3.4in]{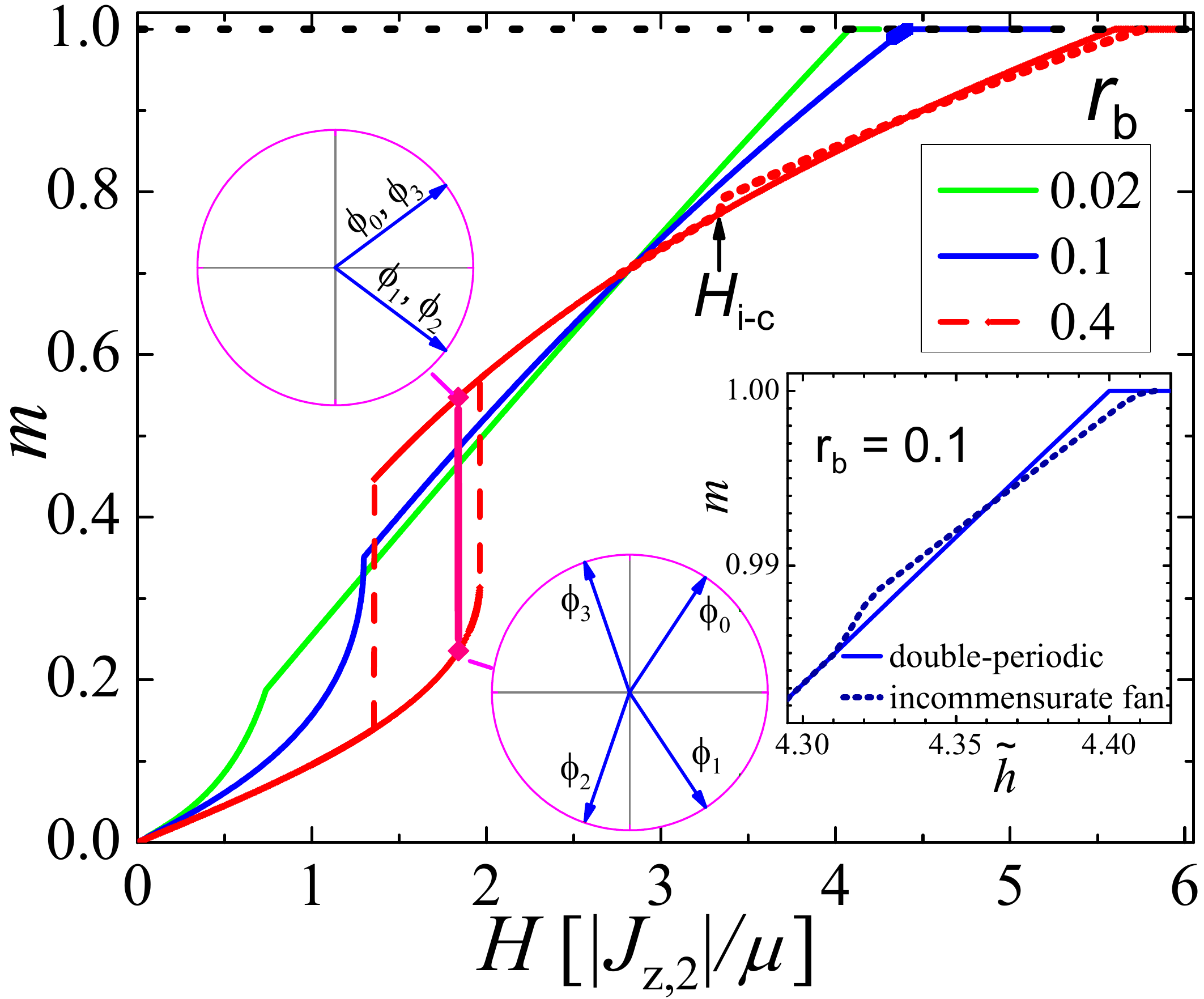}
	\caption{The representative magnetic-field dependences of the reduced magnetization
		for the rotationally-degenerate system for different values of the
		ratio $r_{b}=J_{z,b}/|J_{z,2}|$. The vertical solid line for $r_{b}\!=\!0.4$
		shows the location of the $1^{\mathrm{st}}$-order phase transition and the picture insets show the computed spin configurations at the transition point. The dotted lines for $r_b \!=\!0.4$ and for $r_b \!=\! 0.1$ in the inset show the magnetization for the incommensurate fan state. In the main plot, the range of this state for $r_b \!=\! 0.1$ is marked by bold line. For reference, the solid lines showing the magnetization for the double-periodic state are extended to the high-field region where this state does not minimize the  energy any more.   
		For $r_b=0.02$ the range of incommensurate fan state is invisible in this scale. }
	\label{Fig:m-h-k4eq0}
\end{figure}

Figure \ref{Fig:PhDiagRotDeg} summarizes the magnetic phase diagram
in the $r_{b}$-$\tilde{h}$ plane for the model with rotational degeneracy.
With increasing field, the system transforms from the deformed 45$^\circ$
helix into the double-periodic state at the field monotonically increasing
with the ratio $r_{b}=J_{z,b}/|J_{z,2}|$. At $r_{b}<r_{0}$, Eq.~\eqref{eq:r0},
the transition is second order at the field in Eq.~\eqref{eq:InstField}.
At higher $r_{b}$, the transition becomes first order and the second-order-transition line extends into the dotted instability line. The first-order line is obtained by direct numerical comparison of the
energies in Eqs.~\eqref{eq:EnSymState} and \eqref{eq:EnDPerState}
minimized with respect to the corresponding angles. 

%Resub
Further increase of the magnetic field leads to the transition from the double-periodic into incommensurate fan state. We show two approximate results for the boundary between two states. The red dotted line shows the analytical result in Eq.\ \eqref{eq:i-c-trans} valid for small $r_b$. The solid line with circles is obtained by numerical minimization of the energy in Eq.\ \eqref{eq:EnGenFan} with respect to fan states with different amplitudes and wave vectors. This procedure also provides independent numerical verification that the double-periodic state has the lowest energy in the intermediate field range.
Finally, at even higher magnetic field, the continuous transition to the saturated aligned state takes place. 

Figure \ref{Fig:m-h-k4eq0} shows the representative magnetic-field
dependences of the magnetization for different values of the ratio
$r_{b}$. %
%Resub
The solid lines are obtained for the symmetric states by minimizing the energy in Eq.\ \eqref{eq:EnSymState} with respect to the angles $\alpha$ and $\beta$. The magnetization is then computed from Eq.\ \eqref{eq:MagnSymState}. The dotted curves for $r_b=0.4$ in the main plot and for $r_b=0.1$ in the inset correspond to the incommensurate state. They are computed by minimizing the energy in Eq.\ \eqref{eq:EnGenFan} with respect to the fan amplitude $\vartheta$ and wave vector $q$. The magnetization is evaluated using the derivative of the energy with respect to the magnetic field. %     
We see that the low-field behavior is characterized by upward curvature which becomes more pronounced at smaller $r_{b}$. The transition to the double-periodic
state leads to noticeable features in the magnetization curves.
At small $r_{b}$ the transition is manifested by a kink which becomes more pronounced with increasing $r_{b}$, while at high $r_{b}$, a kink is replaced with a first-order jump. 
With further increase of the field, the transition to the 
%Resub
incommensurate-fan state takes place, which is accompanied by a small increase of the magnetization in comparison with the double-periodic state. With decreasing $r_b$, this increase becomes weaker and the field range for incommensurate state rapidly shrinks. At higher fields the magnetization drops below the double-periodic curve. Finally, the transition to the aligned state takes place at the saturation field, which monotonically increases with $r_{b}$. This transition is manifested as a kink in the magnetization curve. We also note that all curves intersect at one point. This feature occurs to be universal and we will discuss it in detail below. 

\section{Finite four-fold anisotropy\label{sec:FiniteK4}}

In this section, we explore the magnetic phase diagram for finite
in-plane four-fold anisotropy parameter $K_{4}$ using the full energy functional
in Eq.~\eqref{eq:InterLayEnerHK4}. For a state with four-spin periodicity,
this reduced energy per spin, $E_{s}=\mathcal{E}_{\mathrm{lay}}/N|J_{z,2}|$,
can be written as
\begin{align}
E_{s} & =\frac{1}{2}\left[\cos\left(\phi_{2}\!-\!\phi_{0}\right)+\cos\left(\phi_{3}\!-\!\phi_{1}\right)\right]+\frac{k_{4}}{16}\sum_{n=0}^{3}\!\cos\left(4\phi_{n}\right)\nonumber \\
+ & \frac{r_{b}}{4}\left[\cos^{2}\left(\phi_{1}\!-\!\phi_{0}\right)\!+\!\cos^{2}\left(\phi_{2}\!-\!\phi_{1}\right)\right.\nonumber \\
+ & \left.\!\cos^{2}\left(\phi_{3}\!-\!\phi_{2}\right)\!+\!\cos^{2}\left(\phi_{0}\!-\!\phi_{3}\right)\right]\!-\frac{\tilde{h}}{4}\sum_{n=0}^{3}\!\cos\left(\phi_{n}\!-\!\theta\right).\label{eq:EsK4gen}
\end{align}
This energy has the symmetry property 
\begin{equation}
E_{s}(\phi_{n},\theta,k_{4})=E_{s}(\phi_{n}+\frac{\pi}{4},\theta+\frac{\pi}{4},-k_{4}),\label{eq:EnSymProp}
\end{equation}
i.~e., the 45$^\circ$ rotation is equivalent to
sign change of the parameter $k_{4}$. %
%Resub
Similar to the in-plane isotropic case, four-layer periodicity is violated in the vicinity of the saturation field.

%Resub
The model in Eq.\ \eqref{eq:EsK4gen} assumes zero nearest-neighbor exchange constant $J_{z,1}$. As noted in the introduction, finite four-fold anisotropy stabilizes the 90$^\circ$ helix within some range of this constant. This range is evaluated in Appendix \ref{app-rangeJz1}. Finite small $J_{z,1}$ does not significantly alter most results of this section. One key parameter which may be substantially affected by finite $J_{z,1}$ is the wave vector of the incommensurate-fan state near the saturation field, see Eq.\ \eqref{eq:OptQ}. Influence of $J_{z,1}$ on this wave vector is evaluated in Appendix \ref{app-QJz1}.
%Discuss finite Jz1, range of stability of 90 helix state appendix \ref{app-rangeJz1}

The four-fold anisotropy fixes the orientation of the helix in zero magnetic
field. For $k_{4}>0$ and $\tilde{h}=0$, the equilibrium angle configuration
following from the energy in Eq.~\eqref{eq:EsK4gen} is $\phi_{n}\!=\!\pi/4\!\mp\!\pi(n\!-\!1)/2$.
This determines the four easy-axis directions in the $xy$ plane, at $\pm 45^\circ$ and $\pm 135^\circ$.
In this case, the behavior is sensitive to the in-plane orientation
of the magnetic field. In the following sections, we consider 
the phase diagrams for the two symmetric field orientations, 45$^{\circ}$ with respect
to the initial moment direction [$\theta\!=\!0$ in Eq.~\eqref{eq:EsK4gen}]
and along this direction ($\theta\!=\!45^\circ$). 

\subsection{Field angle 45$^{\circ}$ with respect to easy axis}

When the field is oriented at 45$^\circ$ with respect
to equilibrium moment direction, the behavior is qualitatively similar
to the case $k_{4}\!=\!0$ with quantitative modifications of the typical
fields and critical parameters caused by the finite anisotropy. Therefore,
we follow the same route as in Sec.~\ref{sec:RotDegen} and just
revise the results accounting for the finite value of $k_{4}$.  For
the finite anisotropy, the energy of the symmetric state with angles defined
in Eq.~\eqref{eq:SymState} becomes
\begin{align}
 & E_{s}\left(\alpha,\beta\right)=\frac{r_{b}}{2}-\cos\left(\beta\!-\!\alpha\right)\!+\frac{r_{b}\!+\!k_{4}}{8}\left[\cos\left(4\alpha\right)\!+\!\cos\left(4\beta\right)\right]\nonumber \\
 & +\frac{r_{b}}{4}\cos\left[2\left(\beta\!+\!\alpha\right)\right]-\frac{\tilde{h}}{2}\left(\cos\alpha\!-\!\cos\beta\right).\label{eq:EnSymStateK4-2}
\end{align}
This gives equations for the equilibrium angles $\alpha$ and $\beta$
\begin{subequations}
\begin{align}
2\sin\left(\beta\!-\!\alpha\right)&\!+\!k_{4}\sin(4\alpha)\!  +\!r_{b}\left\{\sin(4\alpha)\!+\!\sin\left[2\left(\alpha\!+\!\beta\right)\right]\right\}\ \ \ \nonumber \\
&-\tilde{h}\sin\alpha\!=\!0,\label{eq:alpha-betaK4-1}\\
2\sin\left(\beta\!-\!\alpha\right)&\!-\!k_{4}\sin(4\beta)\!  -\!r_{b}\left\{\sin(4\beta)\!+\!\sin\left[2\left(\alpha\!+\!\beta\right)\right]\right\}\ \ \ \nonumber \\
&-\tilde{h}\sin\beta=\!0.\label{eq:alpha-betaK4-2}
\end{align}
\end{subequations}

First, we consider the small-$\tilde{h}$ expansion of the energy, similar
to Eq.~\eqref{eq:SymEs-h-exp}. The calculation details of the expansion
of the angles and energy are presented in Appendix \ref{App:Small-h-exp45}
and the result for the energy is
\begin{align}
 & E_{s}(\tilde{h})\approx-1-\frac{k_{4}}{4}-\frac{\tilde{h}^{2}}{16\left(1\!+\!r_{b}\!+\!k_{4}\right)}\nonumber \\
- & \frac{1}{512\left(2r_{b}\!+\!k_{4}\right)\!\left(1\!+\!r_{b}\!+\!k_{4}\right)^{2}}
\!\left[1\!+\!\frac{\left(2r_{b}\!+\!k_{4}\right)\!\left(r_{b}\!+\!k_{4}\right)}{\left(1+r_{b}+k_{4}\right)^{2}}\right]\!\tilde{h}^{4}.\label{eq:SymK4Es-h-exp}
\end{align}
Contrary to the case considered in Sec.~\ref{sec:RotDegen}, the
rotational degeneracy of helix is already broken at zero magnetic
field. Helix orientation considered in this subsection is favored
by both the four-fold anisotropy and magnetic field. We will use this
energy expansion later, in the consideration of the helix-rotation
transition for different field orientation. The energy expansion in Eq.\ \eqref{eq:SymK4Es-h-exp} gives the low-field behavior of
the magnetization,
\begin{align}
 & m(\tilde{h})\approx\frac{\tilde{h}}{8\left(1\!+\!r_{b}\!+\!k_{4}\right)}\nonumber \\
+ & \frac{1}{128\left(2r_{b}\!+\!k_{4}\right)\!\left(1\!+\!r_{b}\!+\!k_{4}\right)^{2}}
\!\left[1\!+\!\frac{\left(2r_{b}\!+\!k_{4}\right)\!\left(r_{b}\!+\!k_{4}\right)}{\left(1+r_{b}+k_{4}\right)^{2}}\right]\!\tilde{h}^{3}.\label{eq:magn45degLowH}
\end{align}
We see that the four-fold anisotropy decreases the linear susceptibility. It also reduces the upward curvature.  
Similarly to the rotationally-degenerate case, at sufficiently high
magnetic field the system transfers into the double-periodic state
defined by Eq.~\eqref{eq:DoublePerAngles}. In the next subsection,
we consider the influence of the four-fold anisotropy on this state.

\subsubsection{Double-periodic state}

For the finite four-fold anisotropy, the energy of the double-periodic state for
considered field direction becomes
\begin{align}
E_{s}\left(\alpha,\pi\!-\!\alpha\right) & =\frac{r_{b}}{2}-\frac{k_{4}}{4}+\cos\left(2\alpha\right)+\frac{r_{b}\!+\!k_{4}}{2}\cos^{2}\left(2\alpha\right)\nonumber \\
& -\tilde{h}\cos\alpha\label{eq:EnDPerState45}
\end{align}
yielding the equation for the equilibrium angle
\begin{align}
4\cos\alpha_{0}\left[1\!+\!\left(r_{b}\!+\!k_{4}\right)\cos\left(2\alpha_{0}\right)\right]-\tilde{h} & =0.\label{eq:DoublePerStateK4}
\end{align}
For the double-periodic state, the magnetization would reach saturation at the field $\tilde{h}_{\mathrm{sat}}\!=\!4\left(1\!+\!r_{b}\!+\!k_{4}\right)$.
%Resub
%meaning that the four-fold anisotropy enhances the saturation field for this orientation. 
However, as in the isotropic case, the double-periodic state transforms into the incommensurate fan with increasing magnetic field and therefore this field does not have a direct physical meaning. % 
A peculiar property following from Eq.\ \eqref{eq:DoublePerStateK4} is the presence of the universal
point at the field $\tilde{h}\!=\!2\sqrt{2}$, where $\alpha_{0}\!=\!\pi/4$
and the neighboring moment pairs are orthogonal. The key observation
is that the magnetization at this point $m\!=\!\sqrt{2}/2$ does not depend
on the material's parameters $r_{b}$ and $k_{4}$. 

To evaluate stability of the state, we consider deviation from double
periodicity, $\alpha=\alpha_{0}+\psi$, $\beta=\pi-\alpha_{0}+\psi$. Expansion of the energy in Eq.~\eqref{eq:EnSymStateK4-2} with respect to $\psi$,
\[
\delta E_{s}^{(2)}\left(\psi\right)\!=\!\left\{ 2\left[k_{4}\!-\!2(r_{b}\!+\!k_{4})\cos^{2}\left(2\alpha_{0}\right)\right]\!+\!\frac{\tilde{h}}{2}\cos\alpha_{0}\right\} \psi^{2}
\]
gives the stability condition
\begin{equation}
4\left[k_{4}-2(r_{b}+k_{4})\cos^{2}\left(2\alpha_{0}\right)\right]+\tilde{h}\cos\alpha_{0}>0,\label{eq:StabCondK4}
\end{equation}
which determines the instability field $\tilde{h}_{i}$. Combining
this result with the relation $\tilde{h}\cos\alpha_{0}\!=\!2\left[1\!+\!\cos\left(2\alpha_{0}\right)\right]\left[1\!+\!\left(r_{b}\!+\!k_{4}\right)\cos\left(2\alpha_{0}\right)\right]$
following from Eq.~\eqref{eq:DoublePerStateK4}, we obtain a quadratic
equation for $\cos\left(2\alpha_{0}\right)$
\begin{align*}
 & \left[1\!+\!\cos\left(2\alpha_{0}\right)\right]\left[1\!+\!\left(r_{b}\!+\!k_{4}\right)\cos\left(2\alpha_{0}\right)\right]\\
 & =\!4(r_{b}\!+\!k_{4})\cos^{2}\left(2\alpha_{0}\right)\!-\!2k_{4},
\end{align*}
from which we obtain
\begin{align}
\cos\left(2\alpha_{0}\right) & =-\frac{2(1+2k_{4})}{1+r_{b}+k_{4}+\mathcal{R}},\label{eq:k4-45-cos2a}\\
\mathcal{R} & =\left[\left(1\!+\!r_{b}\!+\!k_{4}\right)^{2}\!+\!12(1\!+\!2k_{4})(r_{b}\!+\!k_{4})\right]^{1/2}\nonumber 
\end{align}
giving
\begin{widetext}
\begin{align*}
\cos\alpha_{0}\!= & \frac{\left[6(1\!+\!2k_{4})(r_{b}\!+\!k_{4})\!+\!\left(1\!+\!r_{b}\!+\!k_{4}\right)\left(r_{b}\!-\!k_{4}\right)\!+\!\left(r_{b}\!-\!k_{4}\right)\mathcal{R}\right]^{1/2}}{1+r_{b}+k_{4}+\mathcal{R}}.
\end{align*}
These results allow us to obtain the instability field for finite anisotropy from Eq.~\eqref{eq:StabCondK4},
\begin{align}
 & \tilde{h}_{i}(r_{b},k_{4})=4\cos\alpha_{0}\left[1+\left(r_{b}+k_{4}\right)\cos\left(2\alpha_{0}\right)\right]\label{eq:k4-45-hi}\\
 & =4\frac{\left[6(1\!+\!2k_{4})(r_{b}\!+\!k_{4})\!+\!\left(r_{b}\!-\!k_{4}\right)\left(1\!+\!r_{b}\!+\!k_{4}\!+\!\mathcal{R}\right)\right]^{1/2}\left[1\!-\!\left(r_{b}\!+\!k_{4}\right)\left(1\!+\!4k_{4}\right)\!+\!\mathcal{R}\right]}{\left[1+r_{b}+k_{4}+\mathcal{R}\right]^{2}}.\nonumber 
\end{align}
\end{widetext}
For $k_4\!=\!0$ this result reproduces Eq.\ \eqref{eq:InstField}. The double-periodic state is stable at $\tilde{h}>\tilde{h}_{i}(r_{b},k_{4})$.
In general, the four-fold anisotropy affects the instability field
in rather complicated way. In the expected case of small parameters,
$r_{b},k_{4}\ll1$, the instability field has a simple asymptotics
\begin{equation}
\tilde{h}_{i}(r_{b},k_{4})\simeq4\sqrt{2r_{b}+k_{4}}.\label{eq:k4-45-hi-Assymp}
\end{equation}
 We see that at small $r_{b}$, the four-fold anisotropy increases
the instability field. However, the accurate analysis shows that this is only
correct for $r_{b}<0.09$. 

The nature of the phase transitions between the deformed 45$^\circ$
helix and double-periodic states is determined by the sign of the
quartic-term coefficient $c_{4}$ in the energy expansion with respect to the perturbation $\psi$, $\delta E_{s}^{(4)}=\frac{1}{4}c_{4}\psi^{4}$.
The calculation of this quartic coefficient described in Appendix
\ref{App:quartic-finitek4-45} yields the result
\begin{align}
c_{4}(r_{b},k_{4})\! & =\!20\left(r_{b}\!+\!k_{4}\right)u^{2}\!-\!10k_{4}\!\nonumber \\
- & \frac{\left(1\!-\!u^{2}\right)\left(1\!+\!15\left(r_{b}\!+\!k_{4}\right)u\right)^{2}}{2\left(u+r_{b}\right)},\label{eq:c4k4}\\
u= & -\cos\left(2\alpha_{0}\right)=\frac{2(1+2k_{4})}{1\!+\!r_{b}\!+\!k_{4}\!+\!\mathcal{R}},\nonumber 
\end{align}
where the parameter $\mathcal{R}$ is defined in Eq.~\eqref{eq:k4-45-cos2a}.
At small $r_{b}$ and $k_{4}$, the quartic coefficient is positive
corresponding to the second-order phase transition. The transition becomes first order with increasing of either
$r_{b}$ or $k_{4}$. The boundary $c_{4}(r_{b},k_{4})=0$ is shown
by the navy line in the Fig.~\ref{Fig:rb-k4-behavior}. It is very
close to the linear dependence, $r_{0}(k_{4})\approx0.099-1.22k_{4}$.
We see that the critical value of $r_{b}$, where the nature of the
transition changes, decreases with increasing $k_{4}$ and for $k_{4}>0.081$
the transition is first order for any $r_{b}$. 

As in the rotationally-isotropic case, the double-periodic state transforms into the incommensurate fan, Eq.\ \eqref{eq:GenFan},  in the vicinity of the saturation field. In the next subsection, we consider the latter state in the case of finite four-fold anisotropy.  

%Resub
\subsubsection{Influence of four-fold anisotropy on incommensurate-fan state near
	saturation \label{subs:IncommensK4}}

In this subsection, we consider modifications of the incommensurate-fan
state caused by the finite four-fold anisotropy term, $K_{4}\left(\cos^{4}\phi_{n}\!+\!\sin^{4}\phi_{n}\!-\frac{3}{4}\right),$
in the energy functional, Eq.~\eqref{eq:InterLayEnerHK4}. In this
case, the fan energy, Eq.~\eqref{eq:EnGenFan} has the additional term
$\frac{1}{N}\sum_{n}\frac{k_{4}}{4}\cos\left[4\vartheta\sin\left(qn+\varphi\right)\right]$.
As a consequence, the quadratic term in the energy expansion with
respect to the fan amplitude $\vartheta$, Eq.~\eqref{eq:FanExpan-a2},
acquires additional $q$-independent contribution
\begin{align}
	a_{2}\left(q\right) & =\!-2k_{4}\!-\!2\sin^{2}q\!-\!2r_{b}(1\!-\!\cos q)\!+\!\frac{\tilde{h}}{2}.
	\label{eq:FanExpanK4-a2}
\end{align}
This means that the four-fold anisotropy just increases the saturation
field,
\begin{equation}
	\tilde{h}_{sat}=(2+r_{b})^{2}+4k_{4}\label{eq:hsatK4-ic}
\end{equation}
but does not change much the overall behavior. In particular, the
energy minimum is still realized at the wave vector in Eq.~\eqref{eq:OptQ}.
However, as mentioned above, finite four-fold anisotropy stabilizes 90$^\circ$ helix for some range of the nearest-neighbor exchange interaction constant $J_{z,1}$ and this constant does affect the wave vector of the incommensurate-fan state, see Appendix \ref{app-QJz1}. Importantly, with finite $J_{z,1}$, the period of this state can be both smaller and larger than four layers.

For the quartic coefficient, we obtain
\begin{equation}
	a_{4}\left(q\right)\!=\!\sin^{4}q\!+\!2r_{b}\left(1\!-\!\cos q\right)^{2}\!+\!4k_{4}\!-\!\frac{\tilde{h}}{16}\label{eq:FanExpanK4-a4}
\end{equation}
for $q\neq\frac{\pi}{2},\pi$ and 
\begin{align}
	a_{4}\left(\frac{\pi}{2},\varphi\right)\! & =\!1\!+\!2r_{b}\!+\!4k_{4}\!-\!\frac{\tilde{h}}{16}\nonumber \\
	+ & \frac{1}{3}\left(1\!-\!2r_{b}\!+\!4k_{4}\!-\frac{\tilde{h}}{16}\right)\cos\left(4\varphi\right).\label{eq:FanExpanK4-a4-90}
\end{align}
From the above quadratic and quartic coefficients in Eqs.\ \eqref{eq:FanExpanK4-a2} and \eqref{eq:FanExpanK4-a4}, we find that the energy of the incommensurate state below the saturation field, Eq.\ \eqref{eq:EnSatIncommens}, is modified as
\begin{align}
	& E_{\mathrm{fan}}\left(q\right)=1+r_{b}+\frac{k_{4}}{4}-\tilde{h}\nonumber \\
	& -\frac{\left[\sin^{2}q+r_{b}\left(1\!-\!\cos q\right)+k_{4}-\frac{\tilde{h}}{4}\right]^{2}}
	{\sin^{4}q+2r_{b}\left(1\!-\!\cos q\right)^{2}+4k_{4}\!-\tilde{h}/16}.\label{eq:EnSatIncommens-1}
\end{align}
For small $r_{b}$, near instability one can again neglect the field
dependence of the optimal wave vector $q$ and use $\cos Q=-\frac{r_{b}}{2}$ from Eq.~\eqref{eq:OptQ}
yielding
\begin{align}
	& E_{\mathrm{fan}}\left(Q\right)\approx1+r_{b}+\frac{k_{4}}{4}-\tilde{h}\nonumber \\
	& -\frac{\left[\left(1+\frac{r_{b}}{2}\right)^{2}+k_{4}-\frac{\tilde{h}}{4}\right]^{2}}{\left(1+\frac{r_{b}}{2}\right)^{4}+4k_{4}-\tilde{h}/16}.
	\label{eq:EnSatK4IncommQ}
\end{align}
On the other hand, for the state with $q=\pi/2$ we have
\begin{align}
	& E_{\mathrm{fan}}\left(\pi/2,\varphi\right)=1+r_{b}+\frac{k_{4}}{4}-\tilde{h}\nonumber \\
	& -\frac{\left(1+r_{b}+\frac{k_{4}}{4}-\frac{\tilde{h}}{4}\right)^{2}}
	{1\!+\!2r_{b}\!+\!4k_{4}\!-\!\frac{\tilde{h}}{16}
		+\frac{1}{3}\left[1\!-\!2r_{b}\!+\!4k_{4}\!-\frac{\tilde{h}}{16}\right]\cos\left(4\varphi\right)}.
	\label{eq:EnSatK4-4}
\end{align}
For $1-2r_{b}+4k_{4}-\tilde{h}/16>0$ or, at the instability field,
$3r_{b}-5k_{4}<1$, minimum energy is realized for the double-periodic
state, $\varphi\!=\!\pi/4$, yielding
\begin{align}
	E_{\mathrm{fan}}\left(\pi/2,\pi/4\right) & =1+r_{b}+\frac{k_{4}}{4}-\tilde{h}.\nonumber \\
	- & \frac{3\left(1+r_{b}+k_{4}-\frac{\tilde{h}}{4}\right)^{2}}{2+8r_{b}+8k_{4}-\tilde{h}/8}.
	\label{eq:EnSatK4-4-45}
\end{align}
Comparing the energies of two states in Eqs.\ \eqref{eq:EnSatK4IncommQ} and \eqref{eq:EnSatK4-4-45}, we can estimate the transition field in the limit $r_b\ll 1$
\begin{align}
	& \tilde{h}_{\mathrm{i-c}}\approx4\left(1\!+\!r_{b}\!+\!k_{4}\right)\nonumber \\
	& -2\left(\sqrt{\frac{3}{2}}\!+\!1\right)\left[1\!+\!4\left(1\!+\!\sqrt{\frac{2}{3}}\right)\frac{r_{b}}{1\!+\!5k_{4}}\right]r_{b}^{2}.
	\label{eq:i-c-transK4}
\end{align}
From this result and the value of the saturation field in Eq.~\eqref{eq:hsatK4-ic},
we can conclude that the field range of the incommensurate state is still proportional to $r_b^2$ and the four-fold anisotropy has only a small influence on this range.

\subsection{Field along easy axis}

\begin{figure}
\includegraphics[width=3.4in]{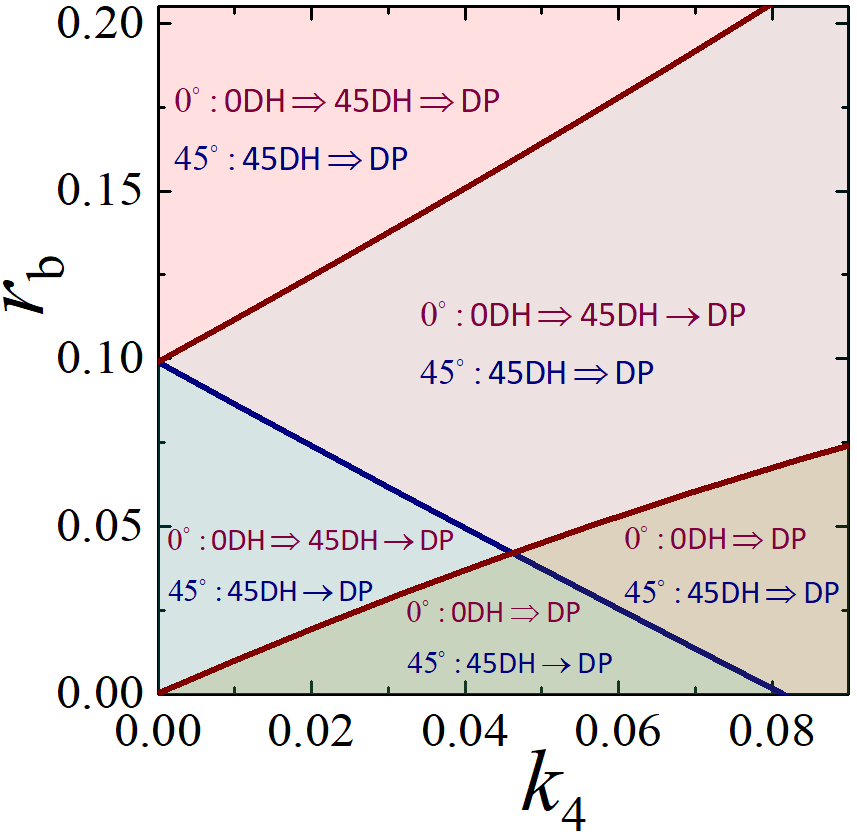}
\caption{The parameter regions with different spin-state-transformations scenarios
for two field orientations. Here 0DH, 45DH, and DP stand for deformed
0$^\circ$ helix, deformed 45$^\circ$
helix, and double-periodic state, respectively, see Fig.~\ref{Fig:SpinStatesK4}.
Single ($\rightarrow$) and double ($\Rightarrow$) arrows indicate
to a second- and first-order phase transitions, respectively. The
upper brown and navy line corresponds to change from second- to first-order
transition to the DP state for the magnetic field applied along the
easy axis and at 45$^\circ$,
respectively.  Above the lower brown line, for the easy-axis field direction, 
the system first has the helix-rotation transition from 0DH to 45DH state and then transfers into the DP state. Below this line, the system jumps directly from 0DH to DP state bypassing the intermediate 45DH state.}
\label{Fig:rb-k4-behavior}
\end{figure}
The case of field along the equilibrium moment direction is richer
and more complicated than the previous cases. The reason is that for such
field direction, the interaction with the field and the four-fold
anisotropy favor different helix orientations. For $\theta=\pi/4$
in Eq.~\eqref{eq:EsK4gen}, it is convenient to utilize the symmetry
property in Eq.~\eqref{eq:EnSymProp} and make substitution $\phi_{n}\rightarrow\phi_{n}+\pi/4$
which transforms the energy to the same form as for $\theta=0$ except for
the sign reverse in the $k_{4}$ term. Therefore, after this substitution,
the energy per spin becomes
\begin{align}
E_{s} & =\frac{1}{2}\left[\cos\left(\phi_{2}\!-\!\phi_{0}\right)+\cos\left(\phi_{3}\!-\!\phi_{1}\right)\right]\!-\frac{k_{4}}{16}\sum_{n=0}^{3}\cos\left(4\phi_{n}\right)\nonumber \\
+ & \frac{r_{b}}{4}\left[\cos^{2}\left(\phi_{1}\!-\!\phi_{0}\right)\!+\!\cos^{2}\left(\phi_{2}\!-\!\phi_{1}\right)\right.\nonumber \\
+\! & \left.\cos^{2}\left(\phi_{3}\!-\!\phi_{2}\right)+\cos^{2}\left(\phi_{0}\!-\!\phi_{3}\right)\right]
\!-\!\frac{\tilde{h}}{4}\sum_{n=0}^{3}\!\cos\left(\phi_{n}\right).
\label{eq:EsK4Along}
\end{align}
At low magnetic fields, the ground state is given by the deformed 0$^\circ$
helix abbreviated as 0DH (lower 3D picture in Fig.~\ref{Fig:SpinStatesK4}), which is
described by the angles $\phi_{0}=0$, $\phi_{3}=-\phi_{1}=\alpha$,
$\phi_{2}=\pi$. Its energy is given by
\begin{align}
E_{s} & =\frac{1}{2}\left[-1+\cos\left(2\alpha\right)\right]-\frac{k_{4}}{8}\left[1+\cos\left(4\alpha\right)\right]\nonumber \\
 & +r_{b}\cos^{2}\alpha-\frac{\tilde{h}}{2}\cos\alpha\label{eq:EsK4Angle0Sym}
\end{align}
and the equilibrium angle $\alpha_0$ is determined by the equation
\begin{equation}
4\cos\alpha_0\left[1+r_{b}-k_{4}\cos\left(2\alpha_0\right)\right]-\tilde{h}=0.\label{eq:EsK4Angle0SymEq}
\end{equation}
Using substitution $u=\cos\alpha_0$, we rewrite these equations as
\begin{subequations}
\begin{align}
	E_{s} & =-1-\frac{k_{4}}{4}+\left(1+r_{b}+k_{4}\right)u^{2}-k_{4}u^{4}-\frac{\tilde{h}}{2}u,\label{eq:EsK4Angle0SymSubst}\\
	& 4u\left[1+r_{b}+k_{4}-2k_{4}u^{2}\right]=\tilde{h}.\label{eq:EsK4Angle0Sym-u}
\end{align}
\end{subequations}
Contrary to isotropic case, the second equation does not have a compact
analytical solution similar to Eq.\ \eqref{eq:AlignAlpha}. We start with the analytic analysis of helix configuration at small magnetic field.

\subsubsection{Small $\tilde{h}$ expansion and first-order helix-rotation transition}

At small magnetic field, we derive from Eq.~\eqref{eq:EsK4Angle0Sym-u} the expansion of the parameter $u$ with respect to $\tilde{h}$,
\begin{align*}
u\! & \approx\frac{\tilde{h}}{4\left(1\!+\!r_{b}\!+\!k_{4}\right)}+\frac{k_{4}\tilde{h}^{3}}{32\left(1\!+\!r_{b}\!+\!k_{4}\right)^{4}}.
\end{align*}
Substituting this result into Eq.~\eqref{eq:EsK4Angle0SymSubst},
we obtain the expansion of the energy
\begin{equation}
	E_{s}(\tilde{h})\!  \approx \!-1\!-\frac{k_{4}}{4}-\frac{\tilde{h}^{2}}{16\left(1\!+\!r_{b}\!+\!k_{4}\right)}
	 -\frac{k_{4}\tilde{h}^{4}}{256\left(1\!+\!r_{b}\!+\! k_{4}\right)^{4}},
	 \label{eq:EsK4Angle0SymExpan}
\end{equation}
which determines the low-field behavior of the magnetization
\begin{equation}
m(\tilde{h})\approx\frac{\tilde{h}}{8\left(1\!+\!r_{b}\!+\!k_{4}\right)}
+\frac{k_{4}\tilde{h}^{3}}{64\left(1\!+r_{b}\!+k_{4}\right)^{4}}.
\label{eq:magn0degLowH}
\end{equation}
Comparing the cubic terms in this result and in Eq.~\eqref{eq:magn45degLowH},
we can conclude that the small-field magnetization for the easy-axis direction
is lower than for the 45$^\circ$ direction. 

As demonstrated in Sec.~\ref{sec:RotDegen}, the magnetic field favors
the 45$^\circ$ helix orientations, which conflicts
with the initial parallel orientation set by the four-fold anisotropy.
Therefore, at small $k_{4}$, a sufficiently strong magnetic field should reorient the
helix. This rotation transition is somewhat similar to the spin-flop
transition in easy-axis antiferromagnets for the magnetic field applied
along the easy axis, see, e.~g., Ref.~\cite{Gurevich1996Book}.
To find the field at which such helix-rotation spin-flop transition takes place,
$\tilde{h}_{r}$, we have to compare the energy of the 0$^\circ$
helix in Eq.~\eqref{eq:EsK4Angle0SymExpan} with the energy of the
45$^\circ$ helix, which in our case can be obtained
from Eq.~\eqref{eq:SymK4Es-h-exp} with the substitution $k_{4}\rightarrow-k_{4}$.
In the limit $k_{4}\ll1$, we can neglect $k_{4}$ in the field-dependent
terms. This gives transition field in the lowest order with respect
to $k_{4}$,
\begin{equation}
\tilde{h}_{r}\simeq 4\left(1+r_{b}\right)\left[\frac{2r_{b}k_{4}}{1+2r_{b}+3r_{b}^{2}}\right]^{1/4}.\label{eq:hrot-Smallk4}
\end{equation}
In particular, $\tilde{h}_{r}\!\simeq\!4\left[2r_{b}k_{4}\right]^{1/4}$
for $r_{b},k_{4}\ll1$. This corresponds to $\mu H_{r}\!\simeq\!4\left[2J_{z,b}K_{4}\right]^{1/4}\!\sqrt{|J_{z,2}|}$
in real units. Keeping $k_{4}$ in the expansion terms, we can derive
a somewhat more accurate result with the next-order term, 
\begin{align}
\tilde{h}_{r}\!\simeq & 4\left(1\!+r_{b}\right)\left(\frac{2r_{b}k_{4}}{1\!+\!2r_{b}\!+\!3r_{b}^{2}}\right)^{1/4}\nonumber \\
\times & \!\left(1\!-\!\sqrt{\frac{2r_{b}k_{4}}{1\!+\!2r_{b}\!+3r_{b}^{2}}}\right).\label{eq:hrot-Intermk4}
\end{align}
When $k_{4}$ is not very small, the precise location of transition
can be found numerically by the energy comparison. With further increase
of the magnetic field, the system again transforms into the double-periodic state.
In addition, the accurate analysis shows that at sufficiently large
$k_{4}$ the intermediate 45DH state may be bypassed and the 0DH state may jump
directly into the DP state.

\subsubsection{Double-periodic and incommensurate-fan states}

\begin{figure}
\includegraphics[width=3.4in]{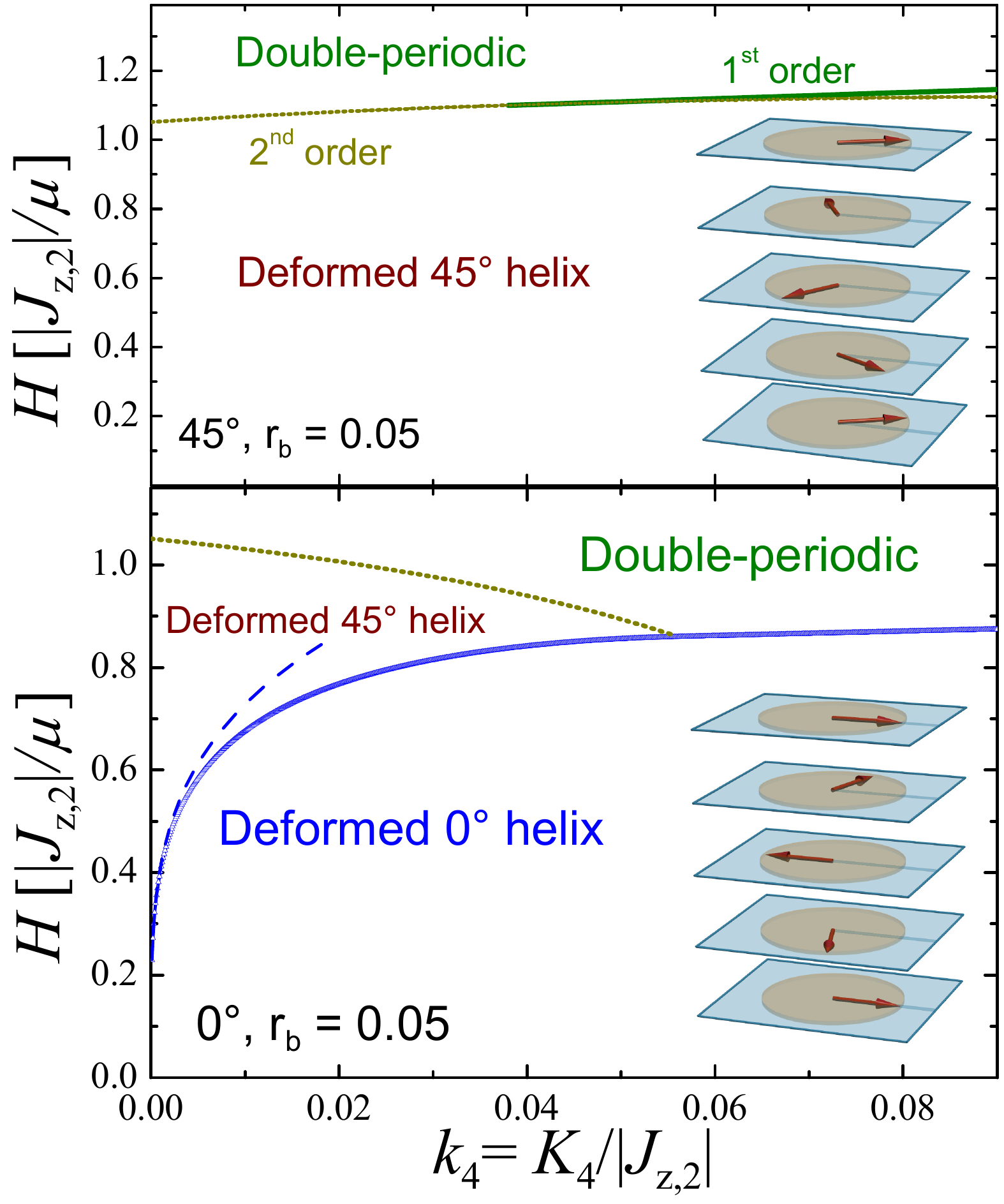}
\caption{Phase diagrams in the plane four-fold anisotropy--magnetic field
for the ratio $r_{b}=0.05$. The upper and lower panel is for the
magnetic field oriented at 45$^\circ$ and
0$^\circ$ with respect to the equilibrium moment direction (easy axis).
In former case, the transition field to the double-periodic state
slowly increases with $k_{4}$ and the transition becomes first order
at $k_{4}>0.0375$. For the parallel direction in the lower plot,
the two subsequent transition 0DH$\Rightarrow$ 45DH$\rightarrow$DP
are realized for $k_{4}<0.055$ and only one first-order transition
0DH$\rightarrow$DP is realized at higher $k_{4}$. The dashed line in
the lower panel presents the low-$k_{4}$ asymptotics of the rotation
transition field $\tilde{h}_{r}$ given by Eq.~\eqref{eq:hrot-Smallk4}. }
\label{Fig:h-k4-rb0_05}
\end{figure}
In the DP state, the energy and the equation for the
equilibrium angle can be obtained from Eq.~\eqref{eq:DoublePerStateK4}
by replacement $k_{4}\rightarrow-k_{4}$,
\begin{subequations}
\begin{align}
	& E_{s}\left(\alpha,\pi-\alpha\right)=\cos\left(2\alpha\right)+\frac{k_{4}}{4}\nonumber \\
	& +\frac{r_{b}}{2}+\frac{r_{b}-k_{4}}{2}\cos^{2}\left(2\alpha\right)-\tilde{h}\cos\alpha,\label{eq:EnDPerStateAlign}\\
	& 4\cos\alpha_{0}\left[1+\left(-k_{4}+r_{b}\right)\cos\left(2\alpha_{0}\right)\right]-\tilde{h}=0.\label{eq:DoublePerStateAlign}
\end{align}
\end{subequations}
%Resub
%This gives the saturation field for the DP state $\tilde{h}_{\mathrm{sat}}\!=\!4\left(1\!-\!k_{4}\!+\!r_{b}\right).$
%It is smaller than for the 45$^\circ$ field orientation, $\tilde{h}_{\mathrm{sat}}(45\lyxmathsym{\textdegree})-\tilde{h}_{\mathrm{sat}}(0\lyxmathsym{\textdegree})=8k_{4}$.
Similarly, the instability field for the double-periodic state $\tilde{h}_{i}$
and the quartic coefficient $c_{4}$ for this field direction can be obtained
from Eqs.~\eqref{eq:k4-45-hi} and \eqref{eq:c4k4} with the same replacement
$k_{4}\!\rightarrow\!-k_{4}$. In particular, at small $r_{b}$ the four-fold anisotropy reduces the instability field for the easy-axis orientation. 
The parameter range where the 45DH $\rightarrow$ DP transition changes
its order is now determined by the condition $c_{4}(r_{b},-k_{4})=0$ using Eq.\ \eqref{eq:c4k4}.
For the easy-axis field direction, the critical value of $r_{b}$ above
which the transition becomes first order increases with $k_{4}$.
This critical value is shown in Fig.~\ref{Fig:rb-k4-behavior} by
the upper brown line. 

At sufficiently large $k_{4}$, two subsequent phase transitions are
replaced by a single first-order phase transition at which the system directly
jumps from the 0DH to DP state. To find the parameter range where
this scenario is realized, we find the field of such direct transition
$\tilde{h}_{t}(r_{b},k_{4})$ by comparing the energies of two states
in Eqs.~\eqref{eq:EsK4Angle0Sym} and \eqref{eq:EnDPerStateAlign}.
The direct-transition scenario is realized when $\tilde{h}_{t}(r_{b},k_{4})>\tilde{h}_{i}(r_{b},-k_{4})$,
where $\tilde{h}_{i}(r_{b},k_{4})$ is given by Eq.~\eqref{eq:k4-45-hi}.
At small $r_b$ and $k_4$ such direct first-order transition takes place for $k_4\!>\!1.04r_b$. 
The lower brown line in Fig.~\ref{Fig:rb-k4-behavior} shows the boundary
below which the direct-transition scenario is realized. 

As mentioned above, the universal point $\alpha_{0}\!=\!\pi/4$ is
realized in the DP state at the field $\tilde{h}\!=\!2\sqrt{2}$, where
magnetization $m\!=\!\sqrt{2}/2$ does not depend on $r_{b}$ and $k_{4}$.
As a consequence, the magnetization is also identical for two field
orientations, meaning that the magnetization curves $m(\tilde{h})$
always cross at this point. 

%Resub
As in other cases, the DP state transforms into the incommensurate-fan state at high fields. The results of subsection \ref{subs:IncommensK4} can be directly applied to the easy-axis orientation using the same substitution $k_{4}\!\rightarrow\!-k_{4}$. In particular, the saturation field for this orientation $\tilde{h}_{sat}=(2+r_{b})^{2}-4k_{4}$
is smaller than for the 45$^\circ$ field orientation in Eq.\ \eqref{eq:hsatK4-ic},
$\tilde{h}_{\mathrm{sat}}(45\lyxmathsym{\textdegree})\!-\!\tilde{h}_{\mathrm{sat}}(0\lyxmathsym{\textdegree})=8k_{4}$.
\begin{figure}
\includegraphics[width=3.4in]{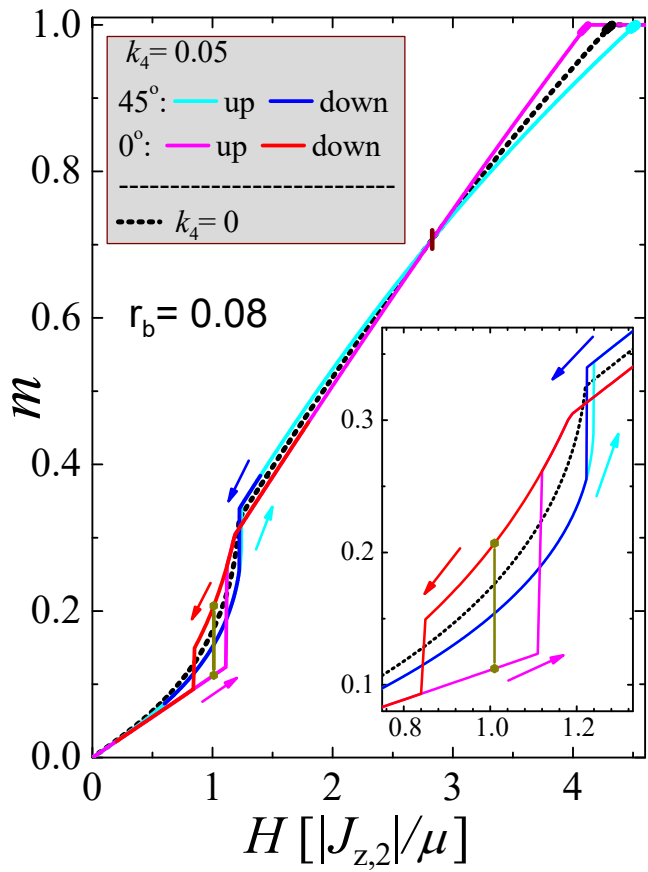}
\caption{The representative field dependences of the magnetization for two
field directions for parameters $r_{b}=0.08$ and $k_{4}=0.05$. The
inset zooms into the transition's regions. For comparison, we also
show by the black dashed line the curve for rotationally-isotropic
case, $k_{4}=0$. For both field orientations the kinks near $\tilde{h}=\mu H/|J_{z,2}|\approx1.2$
correspond to the phase transitions from the deformed $45^\circ$
helix to double-periodic state. For the $0^\circ$
($45^\circ$) orientation the anisotropy
slightly decreased (increases) the transition field and for the $45^\circ$
orientation the transition becomes of the first order. For the parallel
orientation, there is also the first-order phase transition near $\tilde{h}=\mu H/|J_{z,2}|\approx1$
marked by the vertical line corresponding to the helix rotation, as
illustrated in Fig.~\ref{Fig:SpinStatesK4}. All magnetization curves
intersect at the point $\tilde{h}=\mu H/|J_{z,2}|=2\sqrt{2},$ $m=\sqrt{2}/2$
marked by the vertical bar. The bold lines near the saturation show the regions occupied by the incommensurate-fan state.}
\label{Fig:mh-rb0_05k40_02}
\end{figure}

\subsection{Phase diagrams and magnetization curves}

Figure \ref{Fig:h-k4-rb0_05} shows the phase diagrams in the $k_{4}$-$\tilde{h}$
plane for $r_{b}=0.05$ and two field orientations for fields significantly lower than the saturation region. 
The upper and lower panel is for angle 45$^\circ$ and 0$^\circ$
between the magnetic field and the equilibrium moment direction, respectively.
In the former case, the transition field to the double-periodic state
slowly increases with $k_{4}$. The transition becomes first order
at $k_{4}>0.0375$. In the lower panel, the two subsequent transition
0DH$\Rightarrow$ 45DH$\rightarrow$DP are realized at small $k_{4}$
with opposite dependences of the transition fields on $k_{4}$. At $k_{4}=0.055$
the transition lines cross and only one first-order transition 0DH$\rightarrow$DP
is realized at higher $k_{4}$. The dashed line in the lower panel presents
the low-$k_{4}$ asymptotics of the rotation transition field $\tilde{h}_{r}$
given by Eq.~\eqref{eq:hrot-Smallk4}. We can see that it gives an
accurate estimate of the transition field only at very small $k_{4}$,
$k_{4}\lesssim 0.01$.

Figure \ref{Fig:mh-rb0_05k40_02} shows the field dependences of the
magnetization curves for the two field orientations. The plots are made using the representative parameters $r_{b}=0.08$ and $k_{4}=0.05$. %
%Resub
The curve for 45$^\circ$ orientation is obtained in a way similar to $k_4\!=\!0$ case in Fig.\ \ref{Fig:m-h-k4eq0}.
For 0$^\circ$ orientation, we minimized the energy in Eq.\ \eqref{eq:EsK4Along} with respect to four angles $\phi_i$.
In this case, we found that the ground-state configuration always corresponds to one of the symmetric states shown in Fig.\ \ref{Fig:SpinStatesK4}. %
One can observe several key features.
Both field orientations are characterized by the transition to the
DP state near $\tilde{h}\approx1.2$. For the 45$^\circ$
orientation the transition is weak first-order one and is located
at a somewhat higher field than for the parallel orientation. In addition,
for the easy-axis orientation, the first-order helix rotation transition
takes place at a somewhat lower field, near $\tilde{h}\approx1$. The
magnetization curves for the two field orientations cross three times:
at smallest fields, the magnetization is lower for the easy-axis direction,
it becomes larger for this direction after the first-order transition
into the 45DH state and it remains larger until the transition into
the DP state. Note that this small field range is not universal and
vanishes for larger $k_{4}$. In the DP state, the magnetization is
again lower for the easy-axis direction until two magnetization curves
cross at the universal point $\tilde{h}=2\sqrt{2}$. 

%Resub
In both cases, there are narrow incommensurate-fan regions near the saturation. These regions are marked by bold lines in the plot. Small modifications of the magnetization in these regions are similar to one shown in the inset of Fig.\ \ref{Fig:m-h-k4eq0}. %
Finally, the magnetization in the easy-axis direction saturates at a smaller magnetic
field than for the 45$^\circ$ orientation. Qualitatively,
these generic shapes of magnetization curves realize for other sets
of parameters and their features can be used for experimental determination
of the equilibrium helix orientation at zero magnetic field. 

\section{Summary and discussion}
\label{sec:summary}

In summary, we investigated a phenomenological phase diagram for the magnetic helical state with a 90$^{\circ}$ turn angle between neighboring spin layers in the external magnetic field applied perpendicular to the helix axis. We assumed that this unusual state is stabilized by the biquadratic nearest-neighbor interaction and in-plane four-fold anisotropy. 
We found the metamagnetic transition from the distorted helix into the double-periodic state. The corresponding transition field is mostly determined by the strength of the biquadratic interaction. In addition, this field depends on the four-fold anisotropy and field orientation. Depending on the parameters, the transition to the double-periodic state can be either second or first order.  Such behavior is different from the helical structures realized within the frustrated Heisenberg model, where the nature of the transition is fully determined by the modulation wave vector \cite{JohnstonPhysRevB.96.104405}.
When the magnetic field is applied along the equilibrium moment direction (easy axis) and the  four-fold anisotropy $K_4$ is weak, we found an additional first-order spin-flop transition corresponding to the 45$^\circ$ rotation of the distorted helix. The field of this transition behaves as $K_4^{1/4}$ for $K_4\rightarrow 0$. At sufficiently large $K_4$, this helix rotation is bypassed and there is only one  first-order spin-flop transition directly into the double-periodic state.

%Resub
In the vicinity of the saturation field, the double-periodic state transforms into the incommensurate fan. The field range of the latter state is proportional to the biquadratic coupling squared. In the model with rotational degeneracy, the fan period is always smaller than four layers. On the other hand, for the model with the four-fold anisotropy and finite nearest-neighbor exchange constant, the period may be both smaller and larger than four layers. Interestingly,  recent neutron-scattering results \cite{IshidaPNAS21} suggest that, at high magnetic fields, the structure period becomes a little bit larger than four layers corresponding to the wave vector dropping below $\pi/2$.

%Role of fluctuations: transition fields vanish for $T\rightarrow T_m$.
We evaluated the phase diagrams within a mean-field theory completely neglecting thermal spin fluctuations. These fluctuations grow when the temperature approaches the magnetic transition point $T_m$. We expect that the transition magnetic fields computed here will be reduced by the spin fluctuations and vanish as  $T\rightarrow T_m$. The fluctuations also may significantly modify the shapes of magnetization features at the transitions. 

The unusual helical state considered in this paper has been established in the superconducting iron arsenide RbEuFe$_{4}$As$_{4}$\cite{IslamPreprint2019,IidaPhysRevB.100.014506}, which has the superconducting transition at 36.5 K and magnetic transition at 15K. 
The most likely reason for very small nearest-neighbor bilinear exchange interaction $J_{z,1}$ in this material is an accidental compensation of the normal and superconducting RKKY contributions to this parameter \cite{KoshelevPhysRevB19}. In addition, the biquadratic nearest-neighbor term probably has a superconducting origin  
%Resub
due to sensitivity of the superconducting energy to the exchange field. 

Since the interlayer helical state emerges inside the superconducting state, the global magnetic response is hidden either by superconducting screening or by the presence of superconducting vortex lines. This complicates direct verification of the predicted fine features in the magnetization. 
The magnetic field is not uniform in the superconducting state. In the Meissner state, it drops at the scale of the London penetration depth from the surface and in the equilibrium vortex state it oscillates with the periods given by vortex-lattice spacings. 
Even more severe disturbing factor is the formation of the critical state due to vortex pinning in which the magnetic field is macroscopically nonuniform. Due to the temperature dependence of the magnetic susceptibility, such state is formed even for cooling in fixed external magnetic field\cite{VlaskoPhysRevB.99.134503,*VlaskoVlasovPhysRevB.101.104504}.
The distinct features in the global magnetization at the transition fields will be smeared because of these field spatial variations.  

The magnetic field varies at spatial scales much larger than the distance between the Eu$^{2+}$ moments. Therefore, in the simplest scenario, we expect that the local spin configuration and magnetization follow the local magnetic field. 
%Can local probes sense these transitions? 
The saturation field for the in-plane orientation is about 1 kG \cite{Smylie2018}. The metamagnetic transition fields depend on the material's parameters $K_4$ and $J_b$ that are currently unknown. As these parameters are expected to be small in comparison with $J_{z,2}$, it is feasible that the metamagnetic transition fields may be smaller than the in-plane lower critical field $H_{c1}^{ab}$, which for this material is roughly 200 G at low temperatures. In this case, even  in the Meissner state, the magnetic behavior becomes nontrivial:  the spin configuration at the surface will transform with increasing magnetic field and the boundary between two different spin states will be formed parallel to the surface. Similarly, the spin configuration near the center of an isolated in-plane vortex line will be different from the configuration outside and the vortex core will be surrounded by the boundary with the shape of an elliptical cylinder.  Such unusual magnetic vortex structure may have a substantial influence on the properties of the vortex state. In particular, it may be relevant for the understanding of clustering instabilities found in RbEuFe$_{4}$As$_{4}$ by magnetooptical imaging of side faces  \cite{VlaskoPhysRevB.99.134503,*VlaskoVlasovPhysRevB.101.104504}.

\begin{acknowledgments}
I would like to thank V.\ Vlasko-Vlasov for careful reading the manuscript and useful comments.
This work was supported by the US Department of Energy, Office of Science, Basic Energy Sciences, Materials Sciences and Engineering Division. 
\end{acknowledgments}

\appendix
%Resub
\section{Range of stability of 90$^{\circ}$ helix for finite nearest-neighbor
	exchange interaction\label{app-rangeJz1}}

With finite nearest-neighbor coupling, the energy in Eq.~\eqref{eq:InterLayEnerHK4}
at zero magnetic field becomes
\begin{align}
	E_{s} & =\frac{1}{N}\sum_{n}\Big[|J_{z,2}|\cos\left(\phi_{n+2}\!-\!\phi_{n}\right)-J_{z,1}\cos\left(\phi_{n+1}\!-\!\phi_{n}\right)\nonumber \\
	+ & J_{z,b}\cos^{2}\left(\phi_{n+1}-\phi_{n}\right)+\frac{K_{4}}{4}\cos\left(4\phi_{n}\right)\Big].\label{eq:InterLayEnerHK4-Jz1}
\end{align}
We find the range of $J_{z,1}$ within which 90$^{\circ}$ helix still
gives the ground state. Substituting the helix ansatz, $\phi_{n}=qn+\pi/4$,
we obtain
\begin{align}
	E_{s}\left(q\right) & =-|J_{z,2}|+\left(2|J_{z,2}|+J_{z,b}\right)\cos^{2}q\nonumber \\
	& -J_{z,1}\cos q-\frac{K_{4}}{4}\left\langle \cos\left(4qn\right)\right\rangle _{n}.
\end{align}
For the incommensurate state, the four-fold anisotropy vanishes. In
this case, we find the optimal wave vector
\begin{equation}
	\cos Q=\frac{J_{z,1}}{2\left(2|J_{z,2}|\!+\!J_{z,b}\right)}\label{eq:OptWaveVectJ1}
\end{equation}
and the corresponding incommensurate-state energy 
\begin{equation}
	E_{s}\left(Q\right)=-|J_{z,2}|-\frac{J_{z,1}^{2}}{4\left(2|J_{z,2}|\!+\!J_{z,b}\right)}.\label{eq:IncomEnerJz1}
\end{equation}
On the other hand, the energy of 90$^{\circ}$ helix is
\begin{equation}
	E_{s}\left(\pi/2\right)=-|J_{z,2}|\!-\frac{K_{4}}{4}.\label{eq:En90HelixJ1}
\end{equation}
Comparing energies in Eqs.~\eqref{eq:IncomEnerJz1} and \eqref{eq:En90HelixJ1},
we find that the 90$^{\circ}$ helix gives ground state if the condition
\begin{equation}
	|J_{z,1}|<\sqrt{\left(2|J_{z,2}|\!+\!J_{z,b}\right)K_{4}}.\label{eq:Jz1Condition}
\end{equation}
is satisfied. This result is approximate, because we limit ourselves
by a simple incommensurate helical state and did not consider more
complicated nonuniform configurations which may emerge at the transition. 

\section{Calculation of quartic coefficient in the energy expansion for $k_{4}=0$\label{App:quartic-k40}}

Presenting the angles as $\alpha=\alpha_{0}+\psi+\vartheta$, $\beta=\pi-\alpha_{0}+\psi-\vartheta$,
we derive expansion of the energy in Eq.~\eqref{eq:EnSymState} with
respect to $\psi$ and $\vartheta$
\begin{align*}
 & \delta E_{s}\left(\psi,\vartheta\right)\approx\left[-2\cos\left(2\alpha_{0}\right)-2r_{b}\cos\left(4\alpha_{0}\right)+\frac{\tilde{h}}{2}\cos\alpha_{0}\right]\vartheta^{2}\\
 & +\left[\!-4r_{b}\cos^{2}\left(2\alpha_{0}\right)\!+\frac{\tilde{h}}{2}\cos\alpha_{0}\!\right]\\
 & +\!\left(8r_{b}\sin4\alpha_{0}\!-\frac{\tilde{h}}{2}\sin\alpha_{0}\right)\vartheta\psi^{2}\\
 & +\left[\frac{16}{3}r_{b}\cos^{2}\left(2\alpha_{0}\right)-\frac{\tilde{h}}{24}\cos\alpha_{0}\right]\psi^{4}
\end{align*}
Excluding $\vartheta$ for fixed $\psi$
\[
\vartheta=-\frac{8r_{b}\sin(4\alpha_{0})-\frac{\tilde{h}}{2}\sin\alpha_{0}}{-4\cos\left(2\alpha_{0}\right)-4r_{b}\cos\left(4\alpha_{0}\right)+\tilde{h}\cos\alpha_{0}}\psi^{2},
\]
we obtain the full quartic term for expansion with respect to $\psi$
\begin{align*}
E_{s}^{(4)} & =\frac{1}{4}c_{4}\psi^{4},\\
c_{4} & =\frac{64}{3}r_{b}\cos^{2}\left(2\alpha_{0}\right)-\frac{\tilde{h}}{6}\cos(\alpha_{0})\\
- & \frac{\left(8r_{b}\sin4\alpha_{0}-\frac{\tilde{h}}{2}\sin\alpha_{0}\right)^{2}}{-2\cos\left(2\alpha_{0}\right)-2r_{b}\cos\left(4\alpha_{0}\right)+\frac{\tilde{h}}{2}\cos\alpha_{0}}.
\end{align*}
At the instability point, we have relations given by Eqs.~\eqref{eq:DoublePerState}
and \eqref{eq:DoublePerStab}. This allows us to exclude both $\tilde{h}$
and $\alpha_{0}$ and express $c_{4}$ at the instability point via
$r_{b}$ leading to\emph{ }Eq.\emph{~}\eqref{eq:DoublePer-c4} of
the main text.

\section{Calculations for finite four-fold anisotropy and for field angle
45$^{\circ}$ with respect to the moment direction}

\subsection{Small-$\tilde{h}$ expansion\label{App:Small-h-exp45}}

We follow essentially the same steps as in derivation of expansion
in Eq.~\eqref{eq:SymEs-h-exp}. Small magnetic field leads to small
deviations, which we represent as $\alpha\!=\!\frac{\pi}{4}\!-\!\alpha_{+}\!-\frac{\alpha_{-}}{2},\beta\!=\!\frac{\pi}{4}+\!\alpha_{+}\!-\frac{\alpha_{-}}{2}$.
The energy can be expanded as 
\begin{align}
 & E_{s}\left(\alpha_{+},\alpha_{-}\right)\approx\!-1\!-\frac{k_{4}}{4}+\!2\left(1\!+\!r_{b}\!+\!k_{4}\right)\alpha_{+}^{2}\!+\!\left(2r_{b}\!+\!k_{4}\right)\frac{\alpha_{-}^{2}}{2}\nonumber \\
 & -\frac{2\left[1\!+\!4\left(r_{b}\!+\!k_{4}\right)\right]\alpha_{+}^{4}}{3}-4\left(r_{b}\!+\!k_{4}\right)\alpha_{+}^{2}\alpha_{-}^{2}\!-\frac{1}{6}\left(2r_{b}\!+\!k_{4}\right)\alpha_{-}^{4}\nonumber \\
 & -\frac{\sqrt{2}\tilde{h}}{4}\left(2\alpha_{+}-\alpha_{+}\alpha_{-}-\frac{1}{3}\alpha_{+}^{3}-\frac{1}{4}\alpha_{+}\alpha_{-}^{2}\right).\label{eq:EnSymStateK4ExpanPM}
\end{align}
This gives equation for equilibrium $\alpha_{\pm}$ \begin{subequations}
\begin{align}
 & 4\left(1\!+\!r_{b}\!+\!k_{4}\right)\alpha_{+}\!-\frac{8}{3}\left[1\!+\!4\left(r_{b}\!+\!k_{4}\right)\right]\alpha_{+}^{3}\!-\!8\left(r_{b}\!+\!k_{4}\right)\alpha_{+}\alpha_{-}^{2}\nonumber \\
 & -\frac{\sqrt{2}\tilde{h}}{4}\left(2\!-\!\alpha_{-}\!-\!\alpha_{+}^{2}\!-\frac{1}{4}\alpha_{-}^{2}\right)=\!0,\label{eq:EquilK4-alp-p}\\
 & \left(2r_{b}+k_{4}\right)\alpha_{-}-8\left(r_{b}+k_{4}\right)\alpha_{+}^{2}\alpha_{-}-\frac{2}{3}\left(2r_{b}+k_{4}\right)\alpha_{-}^{3}\nonumber \\
 & +\frac{\sqrt{2}\tilde{h}}{4}\left(\alpha_{+}+\frac{1}{2}\alpha_{+}\alpha_{-}\right)=\!0.\label{eq:EquilK4-alp-pm}
\end{align}
\end{subequations}In the expansion with respect to $\tilde{h}$,
\[
\alpha_{\pm}=\sum_{n=1}^{\infty}\alpha_{\pm}^{(n)}\tilde{h}^{n},
\]
only odd $\alpha_{+}^{(n)}$ and even $\alpha_{-}^{(n)}$ are finite,
$\alpha_{+}^{(2k)}=\alpha_{-}^{(2k-1)}=0$. For expansion coefficients,
we obtain from Eqs.~\eqref{eq:EquilK4-alp-p} and \eqref{eq:EquilK4-alp-pm},
\begin{subequations}
\begin{align}
\alpha_{+}^{(1)}  =&\frac{\sqrt{2}}{8\left(1+r_{b}+k_{4}\right)},\label{eq:K4alp-pm-linear}\\
\alpha_{-}^{(2)}= & \!-\frac{\sqrt{2}}{4\left(2r_{b}\!+\!k_{4}\right)}\alpha_{+}^{(1)}\!=\!-\frac{1}{16\left(2r_{b}\!+\!k_{4}\right)\left(1\!+\!r_{b}\!+\!k_{4}\right)},\label{eq:eq:K4alp-pm-2}\\
\alpha_{+}^{(3)}  =&\frac{\sqrt{2}}{512\left(1\!+\!r_{b}+k_{4}\right)^{4}}\nonumber \\
\times & \left\{ \frac{4\left[1\!+\!4\left(r_{b}\!+\!k_{4}\right)\right]}{3}+\frac{\left(2\!+\!k_{4}\right)\left(1\!+\!r_{b}\!+\!k_{4}\right)}{\left(2r_{b}+k_{4}\right)}\right\} .\label{eq:eq:K4alp-pm-3}
\end{align}
\end{subequations}Substituting this expansion into Eq.~\ref{eq:EnSymStateK4ExpanPM},
we derive the energy expansion
\[
E_{s}(\tilde{h})=\sum_{n=1}^{\infty}E_{s}^{(n)}\tilde{h}^{2n}
\]
\begin{subequations}with
\begin{align}
E_{s}^{(2)} & =-\frac{1}{16\left(1+r_{b}+k_{4}\right)}\label{eq:K4En2}\\
E_{s}^{(4)} & =-\frac{1}{512\left(2r_{b}\!+\!k_{4}\right)\left(1\!+\!r_{b}\!+\!k_{4}\right)^{2}}\nonumber \\
\times & \left[1\!+\!\frac{\left(2r_{b}\!+\!k_{4}\right)\left(r_{b}\!+\!k_{4}\right)}{\left(1+r_{b}+k_{4}\right)^{2}}\right].\label{eq:K4En4}
\end{align}
\end{subequations}This result is presented in Eq.~\eqref{eq:SymK4Es-h-exp}
of the main text.

\subsection{Calculation of quartic coefficient in the energy expansion \label{App:quartic-finitek4-45}}

To find the full quartic term, we present $\alpha=\alpha_{0}+\psi+\vartheta$,
$\beta=\pi-\alpha_{0}+\psi-\vartheta$ and expand the energy in Eq.~\eqref{eq:EnSymStateK4-2}
with respect to small deviations $\psi$ and $\vartheta$,
\begin{align*}
E_{s} & \approx\cos\left(2\alpha_{0}\right)+\frac{3r_{b}}{4}+\frac{r_{b}+k_{4}}{4}\cos\left(4\alpha_{0}\right)-\tilde{h}\cos\alpha_{0}\\
+ & \left[-2\cos\left(2\alpha_{0}\right)-2\left(r_{b}+k_{4}\right)\cos\left(4\alpha_{0}\right)+\frac{\tilde{h}}{2}\cos\alpha_{0}\right]\vartheta^{2}\\
+ & \left[-4\left(r_{b}+k_{4}\right)\cos^{2}\left(2\alpha_{0}\right)+2k_{4}+\frac{\tilde{h}}{2}\cos\alpha_{0}\right]\psi^{2}\\
+ & \left(8\left(r_{b}+k_{4}\right)\sin(4\alpha_{0})-\frac{\tilde{h}}{2}\sin\alpha_{0}\right)\vartheta\psi^{2}\\
+ & \left[\frac{16}{3}\left(r_{b}+k_{4}\right)\cos^{2}\left(2\alpha_{0}\right)-\frac{8}{3}k_{4}-\frac{\tilde{h}}{24}\cos\alpha_{0}\right]\psi^{4}.
\end{align*}
Finding the minimal value of $\vartheta$ for fixed $\psi$
\[
\vartheta=-\frac{8\left(r_{b}+k_{4}\right)\sin(4\alpha_{0})-\frac{\tilde{h}}{2}\sin\alpha_{0}}
{-4\cos\left(2\alpha_{0}\right)\!-\!4\left(r_{b}\!+\!k_{4}\right)\cos\left(4\alpha_{0}\right)\!+\!\tilde{h}\cos\alpha_{0}}\psi^{2},
\]
we derive the full quartic term,
\begin{align*}
\delta E_{s}^{(4)} & =\frac{1}{4}c_{4}\psi^{4}\\
c_{4} & =\frac{64}{3}\left(r_{b}\!+\!k_{4}\right)\cos^{2}\left(2\alpha_{0}\right)\!-\frac{32}{3}k_{4}\!-\frac{\tilde{h}}{6}\cos(\alpha_{0})\\
- & \frac{\left(8\left(r_{b}+k_{4}\right)\sin(4\alpha_{0})-\frac{\tilde{h}}{2}\sin\alpha_{0}\right)^{2}}{-2\cos\left(2\alpha_{0}\right)\!-\!2\left(r_{b}\!+\!k_{4}\right)\cos\left(4\alpha_{0}\right)\!+\!\frac{\tilde{h}}{2}\cos\alpha_{0}}.
\end{align*}
At the instability point, we have relations in Eqs.~\eqref{eq:DoublePerStateK4}
and \eqref{eq:StabCondK4}, which allows us to exclude both $\tilde{h}$
and $\alpha_{0}$ and express $c_{4}$ at the instability of the double-periodic
state via $r_{b}$ and $k_{4}$ giving the result in Eq.~\eqref{eq:c4k4}.

%Resub
\section{Incommensurate-fan state for finite four-fold anisotropy and nearest-neighbor
	exchange interaction \label{app-QJz1}}

In this appendix, we consider the influence of the finite nearest-neighbor
exchange interaction on the wave vector of the incommensurate fan,
Eq.~\eqref{eq:GenFan}, emerging near the saturation field. With
finite nearest-neighbor exchange constant $J_{z,1}$ and four-fold
anisotropy $K_{4}$, the reduced fan energy in Eq.~\eqref{eq:EnGenFan}
acquires an additional contribution 
\begin{align*}
&\frac{1}{N}\sum_{n}\Big\{ -j_{z,1}\cos\left[2\vartheta\sin\frac{q}{2}\cos\left(nq\!+\!\frac{q}{2}\!+\!\varphi\right)\right]\\
&+\frac{k_{4}}{4}\cos\left[4\vartheta\sin\left(qn\!+\!\varphi\right)\right]\Big\} 
\end{align*}
with $j_{z,1}\!\equiv\!J_{z,1}/|J_{z,2}|$. Expanding the energy with
respect to the fan amplitude $\vartheta$ near the saturation field,
we obtain for the quadratic coefficient, 
\begin{equation}
	a_{2}\left(q\right)\!=\!-2\sin^{2}q\!+\!\left(-j_{z,1}\!+\!2r_{b}\right)(1\!-\!\cos q)\!-\!2k_{4}\!+\frac{\tilde{h}}{2}\label{eq:FanExpanK4Jz1-a2}
\end{equation}
for $q\neq\pi$. Contrary to the four-fold anisotropy, the nearest-neighbor
exchange interaction influences $q$ dependence of the quadratic coefficient.
The optimal wave vector corresponding to the minimum of $a_{2}\left(q\right)$
is given by
\begin{equation}
	\cos Q=\frac{j_{z,1}-2r_{b}}{4}.\label{eq:OptQJz1}
\end{equation}
The important consequence of this result is that in the case of finite
four-fold anisotropy the $\cos Q$ may be positive meaning that the
period of the incommensurate fan may be longer than four layers. This
may happen if $J_{z,1}$ is positive corresponding to ferromagnetic
interaction and exceeds $2J_{b}$. On the other hand, at zero field,
the 90$^{\circ}$ helix is stable if the condition in Eq.~\eqref{eq:Jz1Condition}
is satisfied. This means that the longer period may realize if the
right-hand side of Eq.~\eqref{eq:Jz1Condition} exceeds $2J_{b}$
giving the condition for the four-fold anisotropy $K_{4}>4J_{b}^{2}/\left(2|J_{z,2}|\!+\!J_{z,b}\right)$.
Evaluating the quadratic coefficient at the optimal wave vector 
\[
a_{2}\left(Q\right)\!=-\frac{\left(4\!-\!j_{z,1}\!+\!2r_{b}\right)^{2}}{8}-\!2k_{4}+\frac{\tilde{h}}{2},
\]
we find that the saturation field
\begin{equation}
	\tilde{h}_{\mathrm{sat}}=\left(2-j_{z,1}/2+r_{b}\right)^{2}+4k_{4}\label{eq:hsatJz1}
\end{equation}
is also shifted by $J_{z,1}$.

\bibliography{Helix90}
 
\end{document}